\documentclass[aps,onecolumn,10pt]{revtex4}
\usepackage{epsfig}
\usepackage{graphicx}
\usepackage{amsmath}
\usepackage{amssymb}
\usepackage{hyperref}
\hypersetup{colorlinks=true,linkcolor=red,citecolor=green,urlcolor=blue}
\numberwithin{equation}{section}
\numberwithin{equation}{section}

\begin{document}
\allowdisplaybreaks
\setcounter{equation}{0}

\title{Determining the normalization of the quantum field theory vacuum, with implications for quantum gravity}

\author{Philip D. Mannheim}
\affiliation{Department of Physics, University of Connecticut, Storrs, CT 06269, USA \\
philip.mannheim@uconn.edu\\ }

\date{May 28 2023}

\begin{abstract}
In a standard quantum field theory the norm $\langle \Omega\vert \Omega\rangle$ of the vacuum state is taken to be finite. In this paper we provide a procedure, based on constructing an equivalent wave mechanics, for determining whether or not $\langle \Omega\vert \Omega\rangle$ actually is finite. We provide an example based on a second-order plus fourth-order scalar field theory, a prototype for quantum gravity, in which it is not. In this example the Minkowski path integral with a real measure diverges though the Euclidean path integral does not. Thus in this example contributions from the Wick rotation contour cannot be ignored. Since $\langle \Omega\vert \Omega\rangle$ is not finite, use of the standard Feynman rules is not valid. And while these rules not only lead to states with negative norm, they in fact lead to states with infinite negative norm. However, if the fields in that theory are continued into the complex plane, we show that then there is a domain in the complex plane known as a Stokes wedge in which one can define an appropriate time-independent, positive and finite inner product, viz. the  $\langle L\vert R\rangle$ overlap of left-eigenstates and right-eigenstates of the Hamiltonian; with the vacuum state then being normalizable, and with there being no states with negative or infinite $\langle L\vert R\rangle$ norm. In this Stokes wedge it is the Euclidean path integral that diverges while the Minkowski path integral does not. The concerns that we raise in this paper only apply to bosons since the matrices associated with their creation and annihilation operators are infinite dimensional. Since the ones  associated with fermions are finite dimensional, the fermion theory vacuum is automatically normalizable. We discuss some general implications of our results for quantum gravity studies, and show that they are relevant to the construction of a consistent, unitary and renormalizable quantum theory of gravity.
 
\end{abstract}

\maketitle

\section{The hidden assumption of quantum field theory}
\label{S1}

Consider a free relativistic neutral scalar field with action 
\begin{eqnarray}
I_{\rm S}=\displaystyle{\int}d^4x \tfrac{1}{2}\left[\partial_{\mu}\phi\partial^{\mu}\phi-m^2\phi^2\right],
\label{1.1}
\end{eqnarray}
and wave equation, Hamiltonian, and equal time commutation relation of the form 
\begin{align}
[\partial_{\mu}\partial^{\mu}+m^2]\phi=0,
\nonumber\\
H=\displaystyle{\int } d^3x \tfrac{1}{2}[\dot{\phi}^2+\bar{\nabla}\phi\cdot \bar{\nabla}\phi+m^2\phi^2],
\nonumber\\
[\phi(\bar{x},t),\dot{\phi}(\bar{x}^{\prime},t)]=i\delta^3(\bar{x}-\bar{x}^{\prime}).
\label{1.2}
\end{align}
With $\omega_k=+(\bar{k}^2+m^2)^{1/2}$ solutions to the  wave equation  obey 
\begin{eqnarray}
\phi(\bar{x},t)=\int \frac{d^3k}{\sqrt{(2\pi)^3 2\omega_k}}[a(\bar{k})e^{-i\omega_k t+i\bar{k}\cdot\bar{x}}+a^{\dagger}(\bar{k})e^{i\omega_k  t-i\bar{k}\cdot\bar{x}}],
\label{1.3}
\end{eqnarray}
 and with $[a(\bar{k}),a^{\dagger}(\bar{k}^{\prime})]=\delta^3(\bar{k}-\bar{k}^{\prime})$ the Hamiltonian is given by 
 \begin{eqnarray}
 H=\frac{1}{2}\int d^3k[\bar{k}^2+m^2]^{1/2} \left[a^{\dagger}(\bar{k})a(\bar{k})+a(\bar{k})a^{\dagger}(\bar{k})\right].
 \label{1.4}
 \end{eqnarray}
Given (\ref{1.4}) we can introduce a no-particle state $\vert \Omega\rangle$ that obeys $a(\bar{k})\vert \Omega \rangle=0$ for each $\bar{k}$, and can identify it as the ground state of $H$. This procedure does not specify the value of $\langle \Omega\vert \Omega\rangle$.

For the theory the associated c-number propagator obeys
\begin{align}
&(\partial_t^2-\bar{\nabla}^2+m^2)D(x)=-\delta^4(x),
\label{1.5}
\end{align}
so that
\begin{align}
& D(x)
=\int \frac{d^4k}{(2\pi)^4}\frac{e^{-ik\cdot x}}{(k^2-m^2+i\epsilon)}.
\label{1.6}
\end{align}
If we identify the  propagator as a vacuum matrix element of q-number fields, viz. 
\begin{eqnarray}
D(x)=-i\langle \Omega\vert T[\phi(x)\phi(0)]\vert\Omega\rangle,
\label{1.7}
\end{eqnarray}
then use of the equal time commutation relation gives 
\begin{align}
&(\partial_t^2-\bar{\nabla}^2+m^2)(-i)\langle \Omega\vert T[\phi(x)\phi(0)]\vert\Omega\rangle=-\langle \Omega \vert\Omega \rangle \delta^4(x). 
\label{1.8}
\end{align}
Comparing with (\ref{1.5}) we see that we can only identify $D(x)$ as the matrix element $-i\langle \Omega\vert T[\phi(x)\phi(0)]\vert\Omega\rangle$ if the vacuum is normalized to one, viz. $\langle \Omega\vert \Omega\rangle=1$. Now if the normalization of the vacuum is finite we of course can always rescale it to one. However, that presupposes that the normalization of the vacuum  is not infinite. We are not aware of any proof in the literature that the normalization of the vacuum is not infinite (either in this particular case or in general), and taking it to be finite is a hidden assumption. So in this paper we shall present a procedure for determining whether the normalization of the vacuum state is finite or infinite. The procedure is based on generalizing to quantum field theory what we know from quantum mechanics. 

This paper is organized as follows. In Sec. \ref{S2} we discuss the quantum-mechanical harmonic oscillator, and in Sec. \ref{S3} we adapt this quantum-mechanical analysis to quantum field theory and present an example, the above second-order-derivative scalar field theory, in which $\langle \Omega\vert \Omega\rangle$ is finite. Then in Secs. \ref{S4}, \ref{S5} and  \ref{S6} we present an example,  a second-order-derivative plus fourth-order-derivative scalar field theory, in which $\langle \Omega\vert \Omega\rangle$ is not finite. In Sec. \ref{S7} we show that by continuing dynamical variables into the complex plane we can construct a second-order-derivative  plus fourth-order-derivative scalar field theory inner product that is finite. In Secs. \ref{S8} and \ref{S12} we  provide an analogous path integral analysis of second-order-derivative plus fourth-order-derivative theories and reach the same conclusions. In Sec. \ref{S9} we discuss the effect of interactions.  In Sec. \ref{S10} we discuss fermion field theories. In Secs. \ref{S11} and \ref{S12} we discuss the relevance of our study to the construction of a consistent, unitary and renormalizable quantum gravity theory. In Sec. \ref{S13} we make some final comments.

To orient the reader to the thrust of this paper we note that in the second-order-derivative  plus fourth-order-derivative neutral scalar field theory discussed below we are interested in a propagator of the generic form
\begin{align}
&D(k)=-\frac{1}{(M_1^2-M_2^2)}\left[\frac{1}{(k^2-M_1^2+i\epsilon)}-\frac{1}{(k^2-M_2^2+i\epsilon)}\right]    
\label{1.9}
\end{align}
that is associated with an action and wave equation of the form
\begin{align}
I_S&=\frac{1}{2}\int d^4x\bigg{[}\partial_{\mu}\partial_{\nu}\phi\partial^{\mu}
\partial^{\nu}\phi-(M_1^2+M_2^2)\partial_{\mu}\phi\partial^{\mu}\phi
+M_1^2M_2^2\phi^2\bigg{]},
\label{1.10}
\end{align}
\begin{align}
&(\partial_t^2-\bar{\nabla}^2+M_1^2)(\partial_t^2-\bar{\nabla}^2+M_2^2)
\phi(x)=0,
\label{1.11}
\end{align}
where $\phi(x)$ is the scalar field. This $D(k)$ propagator possess four distinct features: (i) all the poles in the complex $k_0$ plane are real, so that all energy eigenvalues of  the associated quantum Hamiltonian are real, (ii) the standard Feynman $i\epsilon$ prescription causes positive energies to propagate forward in time and negative energies to propagate backward in time, so that the energy spectrum is bounded from below (no Ostrogradski instability), (iii) at the poles there are both positive and negative residues, to suggest that there are eigenstates of  the associated quantum Hamiltonian that have negative Dirac norm (the overlap of a ket with its Hermitian conjugate bra), and  (iv) all of the residues are finite. While the first two of these features are what one would want of any quantum theory, for them to hold the associated quantum Hamiltonian would have to be a quantum observable that acts on a Hilbert space with some appropriate inner product (one not necessarily the Dirac one)  that is finite, positive and time independent. In such a Hilbert space a quantum operator is a quantum observable when it obeys four conditions: (a) that it is self-adjoint with respect to the appropriate inner product,  (b) that all of its eigenvalues are real, (c) that its eigenvectors are normalizable with respect to the appropriate inner product, and (d) that its eigenvectors form a complete set. (We will see in Sec. \ref{S11} that in the pure fourth-order-derivative $D(k)=-1/k^4$ case this fourth condition has to be relaxed since in the limit in which both $M_1^2$ and $M_2^2$ go to zero the $1/(M_1^2-M_2^2)$ factor in $D(k)$ becomes singular, causing the associated Hamiltonian to become a nondiagonalizable Jordan-block Hamiltonian with an incomplete set of eigenstates, all of which have zero norm, except for the ground state, which has an appropriately chosen norm that is both  finite and positive.) 

Now having a negative Dirac norm is not acceptable of a quantum theory as it leads to nonconservation of probability and thus cannot be associated with any quantum observable.   As we will show using the techniques developed in this paper, this  negative norm problem is avoided since it turns out that the Dirac inner product is not finite. But condition (iv) requires that all residues be finite. Thus the Dirac inner product is not the correct inner product for the problem, and not only for the negative residue sector in (\ref{1.9})  but for the positive residue sector as well. So even if we make the negative norm sector mass very heavy, the resulting  theory associated with the lighter mass positive norm sector would still have an unacceptably infinite Dirac inner product, and could thus not serve as an effective low energy theory.  We thus need a different inner product, one  that is finite, positive and time independent, and in Sec. \ref{S7} we will present the appropriate one. It can be described in two equivalent ways, either as the overlap of a ket with its $CPT$ conjugate bra, [a conjugate that reduces to its $PT$ conjugate ($P$ is parity, $T$ is time reversal)  since $C=1$ ($C$ is charge conjugation) for a neutral scalar field], or the overlap of the right-eigenvector $\vert R\rangle$ of the Hamiltonian with its left-eigenvector $\langle L\vert$. As given in (\ref{1.9}), the propagator is a c-number quantity constructed as the Green's function for a fourth-order differential wave equation. For quantum field theory we need to express this c-number as a c-number matrix element of quantum field operators. For the Dirac inner product  we would use $D(x)=i\langle \Omega\vert T[\phi(x)\phi(0)]\vert\Omega\rangle$. However, this does not work in the second-order-derivative  plus fourth-order-derivative case since in that case $\langle \Omega\vert \Omega\rangle$ is not finite. Rather, we must use $D(x)=i\langle \Omega_L\vert T[\phi(x)\phi(0)]\vert\Omega_R\rangle$ as appropriately adjusted below (by going into the complex plane) so as to obtain a Hilbert space in which the quantum Hamiltonian is self-adjoint and $\langle \Omega_L\vert \Omega_R\rangle$ is finite, positive, and time independent. As we thus see, the key aspect of the study of this paper is the  seemingly innocuous fact that the residues of the poles in $D(k)$ are finite.

In order to obtain a consistent functional variation for the scalar field $I_S$ action that is to lead to the equation of motion given above in (\ref{1.11}), we note that since $I_{\rm S}$ depends not just on the field and its first derivative but also on its second derivative, we must hold both $\phi(x)$ and $\partial_{\mu}\phi(x)$ fixed at the endpoints of the action integral.  Consequently we must treat $\phi(x)$ and $\sigma_{\mu}(x)=\partial_{\mu}\phi(x)$ as independent variables. This will enable us to continue $\phi(x)$ into the complex plane (specifically to $\bar{\phi}(x)=-i\phi(x)$ in the following) without needing to continue $\sigma_{\mu}(x)$ as well, just as will be needed (see Sec. \ref{S7}) in order to make the Hamiltonian be self-adjoint. Analogously, for path integral quantization of the theory we must integrate over independent $\phi(x)$ and $\sigma_{\mu}(x)$ paths (viz. precisely the same set of paths used for the functional variation of $I_S$ in the first place). As we will show in Sec. \ref{S12B}, it is a continuation of $\phi(x)$ into the complex plane while not continuing  $\sigma_{\mu}(x)$ that is  needed in order to make the path integral converge. Thus for the quantum theory $\phi(x)$ and $\sigma_{\mu}(x)$ have to be treated as independent dynamical variables, with $\sigma_{\mu}(x)$ not being the derivative of $\phi(x)$. However, at the stationary minimum we will find from Hamilton's equations of motion  that we can identify the stationary value of $\sigma_{\mu}(x)$ as the $\partial_{\mu}$ derivative of $\phi_{STAT}(x)$. For the path integration we introduce  a complete set of basis functions $f_n(x)$ that vanish at the endpoints of the integral in $I_S$ and are orthogonal to the stationary solution. In terms of the $f_n(x)$  the arbitrary path is of the form  $\bar{\phi}(x)=\bar{\phi}_{STAT}(x)+\sum_n a_n f_n(x)$, $\sigma_{\mu}(x)=\partial_{\mu}\bar{\phi}_{STAT}(x)+\sum_n b_{n, \mu} f_n(x)$, where the $b_{n,\mu}$ are completely independent of the $a_{n}$, with the path integration being over all $a_n$ and $b_{n,\mu}$. Thus with $\partial_{\mu}\bar{\phi}_{STAT}(x)$ being the derivative of $\bar{\phi}_{STAT}(x)$, at the stationary minimum we recover the equation of motion given in (\ref{1.11}), viz. an equation all of whose coefficients are real. In a path integration the stationary path is classical  (viz. the classical path that is followed by the center of a wave packet), while the other paths are due to the spreading of the wave packet. Thus, as discussed further in Sec. \ref{S7E}, we see  that even after the continuation into the complex plane the resulting classical limit is nonetheless completely real, just as will be needed in Secs. \ref{S11} and \ref{S12} in order to be able to construct a consistent second-order-derivative  plus fourth-order-derivative quantum theory of gravity with a real classical gravity limit. 

\section{The quantum-mechanical simple harmonic oscillator} 
\label{S2}

For a simple harmonic oscillator with Hamiltonian $H=\tfrac{1}{2}[p^2+q^2]$ and commutator $[q,p]=i$, there are two sets of bases, the wave function basis and the occupation number space basis. The wave function basis is obtained by setting $p=-i\partial/\partial q$ in $H$ and then solving  the Schr\"odinger wave equation $H\psi(q)=E\psi(q)$. In this way we obtain a ground state with energy $E_0=\tfrac{1}{2}$ and wave function $\psi_0(q)=e^{-q^2/2}$. For occupation number space we set $q=(a+a^{\dagger})/\sqrt{2}$ and $p=i(a^{\dagger}-a)/\sqrt{2}$. This yields $[a,a^{\dagger}]=1$ and $H=a^{\dagger} a+1/2$. We introduce a no-particle state $\vert \Omega \rangle$ that obeys $a\vert \Omega \rangle =0$, with $\vert \Omega \rangle$ being the occupation number space ground state with energy $E_0=\tfrac{1}{2}$.  However, in and of itself this does not fix the norm $\langle \Omega\vert \Omega\rangle$ of the no-particle state or oblige it to be finite.

To fix the $\langle \Omega\vert \Omega\rangle$ norm we need to relate the ground states of the two bases. With $a=(q+ip)/\sqrt{2}$ we set
\begin{align}
\langle q \vert a\vert \Omega \rangle= \frac{1}{\sqrt{2}}\left(q+\frac{\partial}{\partial q}\right)\langle q \vert \Omega \rangle=0,
\label{2.1}
\end{align}
and find that $ \langle q \vert \Omega \rangle=e^{-q^2/2}$. We thus identify $\psi_0(q)=\langle q \vert \Omega \rangle$. We now calculate the standard Dirac norm for vacuum, and obtain  
\begin{align}
\langle \Omega \vert \Omega \rangle=\int_{-\infty}^{\infty}dq \langle \Omega \vert q\rangle\langle q\vert \Omega \rangle=
\int_{-\infty}^{\infty}dq \psi^*_0(q)\psi_0(q)
=\int_{-\infty}^{\infty} dq e^{-q^2}=\sqrt{\pi}.
\label{2.2}
\end{align}
We thus establish that the Dirac norm of the no-particle state is finite. And on setting $\psi_0(q)=e^{-q^2/2}/\pi^{1/4}$ we normalize it to one. That we are able to do this is because we know the form of the wave function $\psi_0(q)$.  

While this  procedure is both straightforward and familiar, it works because both the wave function basis approach and occupation number basis approach have something in common, namely that they are both based on an infinite number of degrees of freedom. For the occupation number basis we can represent the creation and annihilation operators as infinite-dimensional matrices labeled by $\vert \Omega\rangle$, $a^{\dagger}\vert \Omega\rangle$, $a^{\dagger 2}\vert \Omega\rangle$ and so on. For the wave function basis the coordinate $q$ is a continuous variable that varies between $-\infty$ and $\infty$. The two sets of bases are both infinite dimensional, one discrete and the other continuous. The advantage of the continuous basis is that it enables us to express the normalization of the vacuum state as an integral with an infinite range, an integral that is then either finite or infinite.

\section{The quantum field theory oscillator} 
\label{S3}

In the quantum field theory case we do not know the form of the wave function solutions to $H\vert \psi\rangle=E\vert \psi\rangle$, since we cannot realize the canonical commutator given in (\ref{1.2}) as a differential relation. Specifically, we cannot satisfy (\ref{1.2}) by setting $\dot{\phi}(\bar{x},t)$ equal to $-i\partial /\partial \phi(\bar{x},t)$ (though we could introduce a functional derivative $\dot{\phi}(\bar{x},t)=-i\delta /\delta \phi(\bar{x},t)$).

However, we can express the Hamiltonian in terms of creation and annihilation operators.   
So what we can then do is reverse engineer what  we did in the quantum-mechanical case.  We thus introduce 
\begin{align}
a(\bar{k})= \frac{1}{\sqrt{2}}[q(\bar{k})+ip(\bar{k})],\quad a^{\dagger}(\bar{k})= \frac{1}{\sqrt{2}}[q(\bar{k})-ip(\bar{k})],
\label{3.1}
\end{align}
so that 
\begin{align}
&[q(\bar{k}),p(\bar{k}^{\prime})]=i\delta^3(\bar{k}-\bar{k}^{\prime}),\quad H=\frac{1}{2}\int d^3k[\bar{k}^2+m^2]^{1/2} [p^2(\bar{k})+q^2(\bar{k})],
\nonumber\\
&\phi(\bar{x},t)=\frac{1}{\sqrt{2}}\int \frac{d^3k}{\sqrt{(2\pi)^3 2\omega_k}}\left[[q(\bar{k})+ip(\bar{k})]e^{-i\omega_k t+i\bar{k}\cdot\bar{x}}+[q(\bar{k})-ip(\bar{k})]e^{i\omega_k  t-i\bar{k}\cdot\bar{x}}\right].
\label{3.2}
\end{align}

These $q(\bar{k})$ and $p(\bar{k})$ operators do not need to bear any relation to any physical position or momentum operators. Their only role here is to enable us to convert the discrete infinite-dimensional basis associated with each $a(\bar{k})$ and $a^{\dagger}(\bar{k})$ into a convenient continuous one. Specifically, we can realize the $[q(\bar{k}),p(\bar{k}^{\prime})]$ commutator by $p(\bar{k}^{\prime})=-i\partial/\partial q(\bar{k}^{\prime})$, with $H$ then becoming a wave operator. In this way  for each $\bar{k}$ we obtain a solution to the Schr\"odinger equation of the form $\psi(\bar{k})=e^{-q^2(\bar{k})/2}/\pi^{1/4}$. We can define a no-particle vacuum that obeys $a(\bar{k})\vert \Omega\rangle$ for each $\bar{k}$. For each $\bar{k}$ we have
\begin{align}
\langle q(\bar{k})\vert a(\bar{k})\vert \Omega \rangle= \frac{1}{\sqrt{2}}\left[q(\bar{k})+\frac{\partial}{\partial q(\bar{k})}\right]\langle q(\bar{k})\vert \Omega \rangle=0,
\label{3.3}
\end{align}
so that $\langle q(\bar{k})\vert \Omega\rangle=e^{-q^2(\bar{k})/2}/\pi^{1/4}$, and thus
\begin{align}
\langle \Omega\vert \Omega \rangle=\Pi_{\bar{k}}\int d q(\bar{k})\langle \Omega\vert q(\bar{k}) \rangle\langle q(\bar{k})\vert \Omega \rangle
=\Pi_{\bar{k}}\int d q(\bar{k})\frac{e^{-q^2(\bar{k})}}{\pi^{1/2}}=\Pi_{\bar{k}}1=1.
\label{3.4}
\end{align}
Thus the vacuum for the full $H$ obeys $\langle \Omega\vert\Omega\rangle=1$, to thus have a finite normalization. In this way we establish that the vacuum state of the free relativistic scalar field is normalizable.

The general prescription then is to convert the occupation number space Hamiltonian into a product of individual occupation number spaces each with its own $\bar{k}$, and then determine whether the equivalent wave mechanics ground state wave functions constructed this way have a finite normalization in the conventional Schr\"odinger wave mechanics theory sense. If they do, then so does the full  vacuum $\vert \Omega \rangle$ of the full $H$. If on the other hand the equivalent wave mechanics wave functions are not normalizable, then neither is the full $\vert \Omega \rangle$.  Our results for the second-order scalar field theory are also presented in \cite{Mannheim2022}, and in this paper we apply them to the higher-derivative theories that are of interest to quantum gravity, showing that for them the standard Dirac norm $\langle \Omega \vert \Omega \rangle$ is not finite.

Once we are able to show that the vacuum state of the free theory is normalizable, this will remain true in the presence of interactions if the interacting theory is renormalizable. Specifically, if the free theory $D(x)=-i\langle \Omega\vert T[\phi(x)\phi(0)]\vert\Omega\rangle$ is finite, which it will be if the free theory $\langle \Omega\vert \Omega\rangle$ is, then the interacting $D(x)=-i\langle \Omega\vert T[\phi(x)\phi(0)]\vert\Omega\rangle$ propagator will equally be finite after renormalization. Consequently, the renormalized $\langle \Omega\vert \Omega\rangle$ will be finite too. Thus to establish the finiteness of the vacuum normalization of a renormalizable interacting theory, we only need to be able to make a creation and annihilation representation of the free theory. We discuss the role of interactions further in Sec. \ref{S9}. 

As well as providing a procedure for determining whether or not $\langle \Omega\vert \Omega\rangle$ is finite, since the procedure enables is to express the free second-order-derivative Hamiltonian $H$ as an ordinary derivative operator, it does so for interactions as well. Specifically, from (\ref{3.2}) we can write $\phi(\bar{x},t)$ as a derivative operator, viz.
\begin{align}
&\phi(\bar{x},t)=\frac{1}{\sqrt{2}}\int \frac{d^3k}{\sqrt{(2\pi)^3 2\omega_k}}\left[\left[q(\bar{k})+\frac{\partial}{\partial q(\bar{k})}\right]e^{-i\omega_k t+i\bar{k}\cdot\bar{x}}+\left[q(\bar{k})-\frac{\partial}{\partial q(\bar{k})}\right]e^{i\omega_k  t-i\bar{k}\cdot\bar{x}}\right].
\label{3.5}
\end{align}
Thus the insertion of (\ref{3.5}) into an interaction Hamiltonian of the form $H_I=\lambda \int d^3x \phi^4(\bar{x},t)$ enables us to write $H_I$, and thus $H+H_I$,  as a derivative operator. While this procedure enables us to in principle set up the Schr\"odinger problem for $H+H_I$ as a wave mechanics problem, it is still quite a formidable one, just as interacting field theories always have been. However, since it would write the theory in terms of fourth-order derivatives, and since it has an interacting vacuum that is normalized to one and excited states that all have positive norm, it does provide an example of  a theory with higher derivatives that is free of negative-norm ghost states. Thus we can anticipate (and in fact find) that the second-order plus fourth-order quantum field theory that we discuss below in Secs. \ref{S4} - \ref{S7} will be ghost free too.

Since the above analysis is driven by the fact that the dimension of the occupation number space is infinite, the analysis can be carried out for any bosonic field. However, because of the Pauli principle, the occupation number space basis  for fermions of any given $\bar{k}$ is finite dimensional. Thus we have to treat fermions separately, and do so in Sec. \ref{S10}.

\section{Higher-derivative quantum field theories}
\label{S4}

Having presented an example of a theory whose vacuum state is normalizable, we now present an example for which $\langle \Omega\vert \Omega\rangle$ is not finite. The example is based on a second-order-derivative plus fourth-order-derivative neutral scalar field theory with action and equation of motion 
\begin{eqnarray}
I_S&=&\frac{1}{2}\int d^4x\bigg{[}\partial_{\mu}\partial_{\nu}\phi\partial^{\mu}
\partial^{\nu}\phi-(M_1^2+M_2^2)\partial_{\mu}\phi\partial^{\mu}\phi
+M_1^2M_2^2\phi^2\bigg{]},
\nonumber\\
&&(\partial_t^2-\bar{\nabla}^2+M_1^2)(\partial_t^2-\bar{\nabla}^2+M_2^2)
\phi(x)=0,
\label{4.1}
\end{eqnarray}
with ${\rm diag}[\eta_{\mu\nu}]=(1,-1,-1,-1)$.  While we now study this particular model just for illustrative purposes, we note that  it actually arises in quantum gravity studies, and in Secs. \ref{S11} and \ref{S12} we shall explore the implications of this study for quantum gravity.

For (\ref{4.1}) the associated propagator obeys
\begin{align}
&(\partial_t^2-\bar{\nabla}^2+M_1^2)(\partial_t^2-\bar{\nabla}^2+M_2^2)D(x)=-\delta^4(x),
\nonumber\\
&D(x)=-\int \frac{d^4k}{(2\pi)^4}\frac{e^{-ik\cdot x}}{(k^2-M_1^2)(k^2-M_2^2)}
=-\int \frac{d^4k}{(2\pi)^4}\frac{e^{-ik\cdot x}}{(M_1^2-M_2^2)}\left[\frac{1}{(k^2-M_1^2)}-\frac{1}{(k^2-M_2^2)}\right].
\label{4.2}
\end{align}
The energy-momentum tensor $T_{\mu\nu}$, the canonical momenta $\pi^{\mu}$ and $\pi^{\mu\lambda}$, and the equal-time commutators appropriate to the higher-derivative theory are given by \cite{Bender2008b} 
\begin{align}
T_{\mu\nu}&=\pi_{\mu}\phi_{,\nu}+\pi_{\mu}^{~\lambda}\phi_{,\nu,\lambda}-\eta_{\mu\nu}{\cal L},
\nonumber\\ 
 \pi^{\mu}&=\frac{\partial{\cal L}}{\partial \phi_{,\mu}}-\partial_{\lambda
}\left(\frac{\partial {\cal L}}{\partial\phi_{,\mu,\lambda}}\right)=-\partial_{\lambda}\partial^{\mu}\partial^{\lambda}\phi- (M_1^2+M_2^2)\partial^{\mu}\phi,
\nonumber\\
 \pi^{\mu\lambda}&=\frac{\partial {\cal L}}{\partial \phi_{,\mu,\lambda}}=\partial^{\mu}\partial^{\lambda}\phi,
\nonumber\\
T_{00}&=\tfrac{1}{2}\pi_{00}^2+\pi_{0}\dot{\phi}+\tfrac{1}{2}(M_1^2+M_2^2)\dot{
\phi}^2-\tfrac{1}{2}M_1^2M_2^2\phi^2
-\tfrac{1}{2}\pi_{ij}\pi^{ij}+\tfrac{1}{2}(M_1^2+M_2^2)\phi_{,i}\phi^{,i}
\nonumber\\
&=\frac{1}{2}\ddot{\phi}^2-\tfrac{1}{2}(M_1^2+M_2^2)\dot{
\phi}^2-\dddot{\phi}\dot{\phi}-[\partial_i\partial^i\dot{\phi}]\dot{\phi}
-\tfrac{1}{2}M_1^2M_2^2\phi^2
-\tfrac{1}{2}\partial_i\partial_j\phi\partial^i\partial^j\phi+\tfrac{1}{2}(M_1^2+M_2^2)\partial_i\phi\partial^i\phi,
\nonumber\\
&[\phi(\bar{0},t),\dot{\phi}(\bar{x},t)]=0, \qquad[\phi(\bar{0},t),\ddot{\phi}(\bar{x},t)]=0, \qquad [\phi(\bar{0},t),\dddot{\phi}(\bar{x},t])=-i\delta^3(\bar{x}),
\nonumber\\
&[\dot{\phi}(\bar{0},t),\ddot{\phi}(\bar{x},t)]=i\delta^3(\bar{x}), \qquad [\dot{\phi}(\bar{0},t),\dddot{\phi}(\bar{x},t)]=0.
\label{4.3}
\end{align}
With the use of these commutation relations we find that 
\begin{eqnarray}
D(x)=i\langle \Omega\vert T[\phi(x)\phi(0)]\vert\Omega\rangle
\label{4.4}
\end{eqnarray}
indeed satisfies the first equation given in (\ref{4.2}), provided that is  that $\langle \Omega\vert \Omega\rangle=1$ \cite{footnote1}.

To check whether  $\langle \Omega\vert \Omega\rangle$ actually is finite, we need to express the scalar field Hamiltonian $H_S=\int d^3x T_{00}$ in terms of creation and annihilation operators and then construct an equivalent wave mechanics. Given that the solutions to (\ref{4.1}) are plane waves,  we set
\begin{eqnarray}
\phi(\bar{x},t)=\int \frac{d^3k}{(2\pi)^{3/2}}\left[a_1(\bar{k})e^{-i\omega_1 t+i\bar{k}\cdot\bar{x}}+a^{\dagger}_1(\bar{k})e^{i\omega_1  t-i\bar{k}\cdot\bar{x}}+a_2(\bar{k})e^{-i\omega_2 t+i\bar{k}\cdot\bar{x}}+a^{\dagger}_2(\bar{k})e^{i\omega_2  t-i\bar{k}\cdot\bar{x}}\right],
\label{4.5}
\end{eqnarray}
where $\omega_1=+(\bar{k}^2+M_1^2)^{1/2}$, $\omega_2=+(\bar{k}^2+M_2^2)^{1/2}$. Given the commutators in (\ref{4.3}) we obtain 
\begin{eqnarray}
&&[a_1(\bar{k}),a^{\dagger}_{1}(\bar{k}^{\prime})]=[2(M_1^2-M_2^2)(\bar{k}^2+
M_1^2)^{1/2}]^{-1}\delta^3(\bar{k}-\bar{k}^{\prime}),
\nonumber\\
&&[a_2(\bar{k}),a^{\dagger}_{2}(\bar{k}^{\prime})]=-[2(M_1^2-M_2^2)(\bar{k}^2+
M_2^2)^{1/2}]^{-1}\delta^3(\bar{k}-\bar{k}^{\prime}),
\nonumber\\
&&[a_1(\bar{k}),a_{2}(\bar{k}^{\prime})]=0,\quad[a_1(\bar{k}),a^{\dagger}_{2}(\bar{k}^{\prime})]=0,\quad[a^{\dagger}_1(\bar{k}),a_{2}(\bar{k}^{\prime})]=0,\quad
[a^{\dagger}_1(\bar{k}),a^{\dagger}_{2}(\bar{k}^{\prime})]=0,
\label{4.6}
\end{eqnarray}
with the Hamiltonian then taking the form 
\begin{eqnarray}
H_S&=&\frac{1}{2}\int d^3k\bigg{[}2(M_1^2-M_2^2)(\bar{k}^2+M_1^2)\left[a^{\dagger}_{1}(\bar{k})a_1(\bar{k})
+a_{1}(\bar{k})a^{\dagger}_1(\bar{k})\right]
\nonumber\\
&-&2(M_1^2-M_2^2)(\bar{k}^2+M_2^2)\left[a^{\dagger}_2(\bar{k})a_{2}(\bar{k})
+a_{2}(\bar{k})a^{\dagger}_2(\bar{k})\right]\bigg{]}
\nonumber\\
&=&\int d^3k\bigg{[}2(M_1^2-M_2^2)(\bar{k}^2+M_1^2)a^{\dagger}_{1}(\bar{k})a_1(\bar{k})
-2(M_1^2-M_2^2)(\bar{k}^2+M_2^2)a^{\dagger}_2(\bar{k})a_{2}(\bar{k})
\nonumber\\
&+&\frac{1}{2}(\bar{k}^2+M_1^2)^{1/2}\delta^3(\bar{0})+\frac{1}{2}(\bar{k}^2+M_2^2)^{1/2}\delta^3(\bar{0})\bigg{]},
\label{4.7}
\end{eqnarray}
where $(2\pi)^3\delta^3(\bar{0})$ is a quantization box volume $V$.
We note that with $M_1^2-M_2^2>0$ for definitiveness, we see negative signs in both $H_S$ and the $[a_2(\bar{k}),a^{\dagger}_{2}(\bar{k}^{\prime})]$ commutator, while noting that despite this the zero-point energy is positive.  We shall see below that the negative sign concerns will be resolved once we settle the issue of the normalization of the vacuum. To do that we now descend to the quantum-mechanical limit of the theory, the Pais-Uhlenbeck oscillator model.

\section{Higher-derivative quantum mechanics} 
\label{S5}

In order to study the Pauli-Villars regulator, in \cite{Pais1950} Pais and Uhlenbeck (${\rm PU}$) introduced a fourth-order quantum-mechanical oscillator model with action and equation of motion
\begin{eqnarray}
I_{\rm PU}=\frac{1}{2}\int dt\left[{\ddot z}^2-\left(\omega_1^2
+\omega_2^2\right){\dot z}^2+\omega_1^2\omega_2^2z^2\right],\qquad \ddddot{z}+(\omega_1^2+\omega_2^2)\ddot{z}+\omega_1^2\omega_2^2z^2=0,
\label{5.1}
\end{eqnarray}
where for definitiveness in the following we take $\omega_1>\omega_2$. As constructed this action possesses three variables $z$, $\dot{z}$ and $\ddot{z}$. This is too many for one oscillator but not enough for two. The system is thus a constrained system. And so we introduce a new variable $x=\dot{z}$ and its conjugate $p_x$. And using the method of Dirac constraints obtain the time-independent Hamiltonian \cite{Mannheim2000,Mannheim2005}
\begin{eqnarray}
H_{\rm PU}=\frac{p_x^2(t)}{2}+p_z(t)x(t)+\frac{1}{2}\left(\omega_1^2+\omega_2^2 \right)x^2(t)-\frac{1}{2}\omega_1^2\omega_2^2z^2(t),
\label{5.2}
\end{eqnarray}
with two sets of canonical equal-time commutators of the form 
\begin{align}
[z(t),p_z(t)]=i, \qquad [x(t),p_x(t)]=i.
\label{5.3}
\end{align}
The form given for $H_{\rm PU}$ can be understood by writing the Hamiltonian as the Legendre transform of the Lagrangian. Specifically, on recalling that $\dot{z}=x$ we obtain  $H_{\rm PU}=p_x(t)\dot{x}(t)+p_z(t)x(t)-(1/2)[\dot{x}^2(t)-(\omega_1^2+\omega_2^2)x^2(t)+\omega_1^2\omega_2^2z^2(t)]$, from which (\ref{5.2}) follows when $\dot{x}(t)=p_x(t)$.

The terms that appear in $H_{\rm PU}$ are in complete parallel to the first four terms in the field theory $T_{00}$ given in (\ref{4.3}), with the PU oscillator model being the nonrelativistic limit of the relativistic scalar field theory, with the spatial dependence having been frozen out. Since canonical commutators only involve time derivatives, freezing out the spatial dependence will still give the full dynamical content of the relativistic theory. In fact we can set $i=[z,p_z]\equiv [\phi,\pi_0]=[\phi,-\dddot{\phi}-(M_1^2+M_2^2)\dot{\phi}]=i\delta^3(\bar{x})$, to thus parallel the commutators given in (\ref{4.3}).    

On setting $p_z=-i\partial_z$, $p_x=-i\partial_x$ the Schr\"odinger problem for $H_{\rm PU}$ can be solved analytically, with
the state with energy $(\omega_1+\omega_2)/2$ having a wave function that is of the form \cite{Mannheim2007} 
\begin{align}
\psi_0(z,x)=\exp[\tfrac{1}{2}(\omega_1+\omega_2)\omega_1\omega_2z^2+i\omega_1\omega_2zx-\tfrac{1}{2}(\omega_1+\omega_2)x^2].
\label{5.4}
\end{align}
While this wave function is well behaved at large $x$, it diverges at large $z$, and consequently as a wave function it  is not normalizable. That the wave function is not normalizable has also been pointed out in \cite{Woodard1989,Woodard2007,Woodard2015}.

To relate this wave function to the no-particle vacuum $\vert \Omega \rangle$ we second quantize the theory. And with the wave equation given in (\ref{5.1}), and with $\dot{z}=i[H_{\rm PU},z]=x$, $\dot{x}=p_x$, $\dot{p}_x=-p_z-(\omega_1^2+\omega_2^2)x$, $\dot{p}_z=\omega_1^2\omega_2^2z$, we obtain 
\begin{eqnarray}
z(t)&=&a_1e^{-i\omega_1t}+a_1^{\dagger}e^{i\omega_1t}+a_2e^{-i\omega_2t}+a_2^{\dagger}e^{i\omega_2t},
\nonumber\\
p_z(t)&=&i\omega_1\omega_2^2
[a_1e^{-i\omega_1t}-a_1^{\dagger}e^{i\omega_1t}]+i\omega_1^2\omega_2[a_2e^{-i\omega_2t}-a_2^{\dagger}e^{i\omega_2t}],
\nonumber\\
x(t)&=&-i\omega_1[a_1e^{-i\omega_1t}-a_1^{\dagger}e^{i\omega_1t}]-i\omega_2[a_2e^{-i\omega_2t}-a_2^{\dagger}e^{i\omega_2t}],
\nonumber\\
p_x(t)&=&-\omega_1^2 [a_1e^{-i\omega_1t}+a_1^{\dagger}e^{i\omega_1t}]-\omega_2^2[a_2e^{-i\omega_2t}+a_2^{\dagger}e^{i\omega_2t}],
\label{5.5}
\end{eqnarray}
and a Hamiltonian and commutator algebra of the form \cite{Mannheim2000}
\begin{align}
H_{\rm PU}&=2(\omega_1^2-\omega_2^2)(\omega_1^2 a_1^{\dagger}
a_1-\omega_2^2a_2^{\dagger} a_2)
+\tfrac{1}{2}(\omega_1+\omega_2),
\label{5.6}
\end{align}
\begin{align}
[a_1,a_1^{\dagger}]&=\frac{1}{2\omega_1(\omega_1^2-\omega_2^2)}, \qquad
[a_2,a_2^{\dagger}]=-\frac{1}{2\omega_2(\omega_1^2-\omega_2^2)}.
\label{5.7}
\end{align}
We note the similarity to (\ref{4.7}) and (\ref{4.6}). 

As constructed, (\ref{5.6}) and (\ref{5.7}) admit of two inequivalent realizations as one can take $a_1$ and either $a_2$ or $a_2^{\dagger}$ to annihilate the vacuum. Thus one can define a Hilbert space in which $a_1\vert\Omega\rangle=0$, $a_2\vert\Omega\rangle=0$, or one can define a separate and distinct Hilbert space in which $a_1\vert\Omega\rangle=0$, $a_2^{\dagger}\vert\Omega\rangle=0$. In the Hilbert space in which $a_1\vert\Omega\rangle=0$, $a_2\vert\Omega\rangle=0$, we note that even though the $a_2^{\dagger} a_2$ term appears in $H_{\rm PU}$ with a minus sign, there is a compensating minus sign in the $[a_2,a_2^{\dagger}]$ commutator. In consequence, in this realization  all energy eigenvalues of $H_{\rm PU}$ are positive, with the no-particle state $\vert \Omega \rangle$ that both $a_1$ and $a_2$ annihilate being the state of lowest energy. However, in this Hilbert space the matrix element  $\langle \Omega\vert a_2a_2^{\dagger}\vert \Omega\rangle$ is negative, the ghost problem of higher-derivative theories. This problem has been solved in the literature \cite{Bender2008a,Bender2008b},  and we shall return to it in detail below.

Alternatively,  if one takes $a_2^{\dagger}\equiv b_2$ to annihilate the vacuum, we obtain a commutator  $[b_2,b_2^{\dagger}]=1/[2\omega_1(\omega_1^2-\omega_2^2)]$ that is positive. However then the energy spectrum of the Hamiltonian becomes unbounded from below, the familiar Ostrogradski instability of higher-derivative theories.   In this case the no-particle state has energy $(\omega_1-\omega_2)/2$ (it is not the lowest lying level in this case), with a wave function $\psi_0(z,x)=\exp[-\tfrac{1}{2}(\omega_1-\omega_2)\omega_1\omega_2z^2-i\omega_1\omega_2zx-\tfrac{1}{2}(\omega_1-\omega_2)x^2]$ \cite{Bender2008a} that with $\omega_1>\omega_2$ is normalizable.

While we thus have to deal with a negative-norm problem or a negative energy problem, we note that the two problems do not occur in one and the same Hilbert space. Thus in any given Hilbert space we at most only have to deal with one. And since an energy spectrum that is unbounded from below is not physical, we shall work solely in the Hilbert space in which $a_1\vert\Omega\rangle=0$, $a_2\vert\Omega\rangle=0$. In Sec. \ref{S12}, which provides a quick explanation of our results,  we shall show that the $a_1\vert\Omega\rangle=0$, $a_2\vert\Omega\rangle=0$ and $a_1\vert\Omega\rangle=0$, $a_2^{\dagger}\vert\Omega\rangle=0$ realizations correspond to different and thus inequivalent Feynman $i\epsilon$ prescriptions.

In the Hilbert space in which both $a_1$ and $a_2$ annihilate the vacuum the energy spectrum is bounded from below, and the energy of the ground state is $(\omega_1+\omega_2)/2$. On solving the time-dependent Schr\"odinger equation the wave function of the ground state is $\psi_0(z,x)e^{-i(\omega_1+\omega_2)t/2}$, where $\psi_0(z,x)$ is given in (\ref{5.4}). For this wave function the normalization of $\vert \Omega \rangle$ is then  given by 
\begin{align}
\langle \Omega\vert \Omega\rangle=\int_{-\infty}^{\infty} dz\int_{-\infty}^{\infty}dx\langle \Omega\vert z,x\rangle\langle z,x\vert\Omega\rangle=\int_{-\infty}^{\infty} dz \int_{-\infty}^{\infty}dx \psi_0^*(z,x)\psi_0(z,x).
\label{5.8}
\end{align}
With $\psi_0(z,x)$ diverging at large $z$, this normalization integral is infinite. Thus we see that through our knowledge of the form of the ground state wave function as given in (\ref{5.4}) we are able to determine the normalization of the PU theory vacuum and  establish that it is infinite. We can thus anticipate and will immediately show in Sec. \ref{S6} that this is also the case for the second-order plus fourth-order scalar quantum field  theory as well. Then in Sec. \ref{S7} we will discuss what to do about it, with there actually being a mechanism for obtaining a finite normalization \cite{Bender2008a,Bender2008b}, one that also takes care of the fact that according to (\ref{5.7}) $\langle \Omega\vert a_2a_2^{\dagger}\vert\Omega\rangle$ is negative.

Already in \cite{Mannheim2007}  it was known that for the ground state of the second-order plus fourth-order quantum-mechanical PU oscillator model the vacuum $\langle \Omega\vert \Omega\rangle$ norm was infinite. However, it was not known whether the vacuum $\langle \Omega\vert \Omega\rangle$ norm of the second-order plus fourth-order quantum field theory was finite or infinite. Using the procedure developed in this paper we can now determine whether it is finite or infinite, and in Sec. \ref{S6} we show that $\langle \Omega\vert \Omega\rangle$ is in fact infinite. Then in Sec. \ref{S7} we show that there is another inner product, viz. the 
$\langle L\vert R\rangle$ overlap of left-eigenstates and right-eigenstates of the Hamiltonian,  that is finite.

\section{The nonnormalizable vacuum of higher-derivative field theories}
\label{S6}
 To determine the second-order plus fourth-order scalar field theory vacuum normalization we first need to invert (\ref{5.5}). This yields
\begin{align}
a_1e^{-i\omega_1t}&=\frac{1}{2(\omega_1^2-\omega_2^2)}\left[-\omega_2^2z(t)-p_x(t)+i\omega_1x(t)+i\frac{p_z(t)}{\omega_1}\right], 
\nonumber\\
a_1^{\dagger}e^{i\omega_1t}&=\frac{1}{2(\omega_1^2-\omega_2^2)}\left[-\omega_2^2z(t)-p_x(t)-i\omega_1x(t)-i\frac{p_z(t)}{\omega_1}\right],
\nonumber\\
a_2e^{-i\omega_2t}&=\frac{1}{2(\omega_1^2-\omega_2^2)}\left[\omega_1^2z(t)+p_x(t)-i\omega_2x(t)-i\frac{p_z(t)}{\omega_2}\right],\
\nonumber\\
a_2^{\dagger}e^{i\omega_2t}&=\frac{1}{2(\omega_1^2-\omega_2^2)}\left[\omega_1^2z(t)+p_x(t)+i\omega_2x(t)+i\frac{p_z(t)}{\omega_2}\right].
\label{6.1}
\end{align}
On generalizing to each $\bar{k}$ and setting $\omega_1(\bar{k}) =+(\bar{k}^2+M_1^2)^{1/2}$, $\omega_2(\bar{k}) =+(\bar{k}^2+M_2^2)^{1/2}$, we obtain
\begin{align}
a_1(\bar{k})e^{-i\omega_1(\bar{k})t}&=\frac{1}{2(M_1^2-M_2^2)}\left[-\omega_2^2(\bar{k})z(\bar{k},t)-p_x(\bar{k},t)+i\omega_1(\bar{k})x(\bar{k},t)+i\frac{p_z(\bar{k},t)}{\omega_1(\bar{k})}\right],
\nonumber\\
a_1^{\dagger}(\bar{k})e^{i\omega_1(\bar{k})t}&=\frac{1}{2(M_1^2-M_2^2)}\left[-\omega_2^2(\bar{k})z(\bar{k},t)-p_x(\bar{k},t)-i\omega_1(\bar{k})x(\bar{k},t)-i\frac{p_z(\bar{k},t)}{\omega_1(\bar{k})}\right],
\nonumber\\
a_2(\bar{k})e^{-i\omega_2(\bar{k})t}&=\frac{1}{2(M_1^2-M_2^2)}\left[\omega_1^2(\bar{k})z(\bar{k},t)+p_x(\bar{k},t)-i\omega_2(\bar{k})x(\bar{k},t)-i\frac{p_z(\bar{k},t)}{\omega_2(\bar{k})}\right],
\nonumber\\
a_2^{\dagger}(\bar{k})e^{i\omega_2(\bar{k})t}&=\frac{1}{2(M_1^2-M_2^2)}\left[\omega_1^2(\bar{k})z(\bar{k},t)+p_x(\bar{k},t)+i\omega_2(\bar{k})x(\bar{k},t)+i\frac{p_z(\bar{k},t)}{\omega_2(\bar{k})}\right].
\label{6.2}
\end{align}
Inverting (\ref{6.2}) gives
\begin{align}
z(\bar{k},t)&=a_1(\bar{k})e^{-i\omega_1(\bar{k})t}+a_1^{\dagger}(\bar{k})e^{i\omega_1(\bar{k})t}+a_2(\bar{k})e^{-i\omega_2(\bar{k})t}+a_2^{\dagger}(\bar{k})^{i\omega_2(\bar{k})t},
\nonumber\\
p_z(\bar{k},t)&=i\omega_1(\bar{k})\omega_2^2(\bar{k})
[a_1(\bar{k})e^{-i\omega_1(\bar{k})t}-a_1^{\dagger}(\bar{k})e^{i\omega_1(\bar{k})t}]+i\omega_1^2(\bar{k})\omega_2(\bar{k})[a_2(\bar{k})e^{-i\omega_2(\bar{k})t}-a_2^{\dagger}(\bar{k})e^{i\omega_2(\bar{k})t}],
\nonumber\\
x(\bar{k},t)&=-i\omega_1(\bar{k})[a_1(\bar{k})e^{-i\omega_1(\bar{k})t}-a_1^{\dagger}(\bar{k})e^{i\omega_1(\bar{k})t}]-i\omega_2(\bar{k})[a_2(\bar{k})e^{-i\omega_2(\bar{k})t}-a_2^{\dagger}(\bar{k})^{i\omega_2(\bar{k})t}],
\nonumber\\
p_x(\bar{k},t)&=-\omega_1^2(\bar{k})[a_1(\bar{k})e^{-i\omega_1(\bar{k})t}+a_1^{\dagger}(\bar{k})e^{i\omega_1(\bar{k})t}]-\omega_2^2(\bar{k})[a_2(\bar{k})e^{-i\omega_2(\bar{k})t}+a_2^{\dagger}(\bar{k})^{i\omega_2(\bar{k})t}].
\label{6.3}
\end{align}
From (\ref{6.3}) and the commutation relations given in (\ref{4.6}) it follows that 
\begin{eqnarray}
&&[z(\bar{k},t),p_z(\bar{k}^{\prime},t)]=\delta^3(\bar{k}-\bar{k}^{\prime}),\qquad [x(\bar{k},t),p_x(\bar{k}^{\prime},t)]=\delta^3(\bar{k}-\bar{k}^{\prime}),
\nonumber\\
&&[z(\bar{k},t),x(\bar{k}^{\prime},t)]=0,\quad[z(\bar{k},t),p_x(\bar{k}^{\prime},t)]=0,\quad[p_z(\bar{k},t),x(\bar{k}^{\prime},t)]=0,\quad
[p_z(\bar{k},t),p_x(\bar{k}^{\prime},t)]=0.
\label{6.4}
\end{eqnarray}
Insertion of (\ref{6.2}) into the Hamiltonian given in (\ref{4.7}) then yields an equivalent, time-independent Hamiltonian
\begin{eqnarray}
H_S&=&\int d^3k\bigg{[}\frac{p_x^2(\bar{k},t)}{2}+p_z(\bar{k},t)x(\bar{k},t)+\frac{1}{2}\left[\omega_1^2(\bar{k})+\omega_2^2(\bar{k}) \right]x^2(\bar{k},t)-\frac{1}{2}\omega_1^2(\bar{k})\omega_2^2(\bar{k})z^2(\bar{k},t)\bigg{]}.
\label{6.5}
\end{eqnarray}
For each momentum state we recognize the quantum field theory Hamiltonian $H_S$ given in (\ref{6.5}) as being of precisely the form of the quantum-mechanical $H_{\rm PU}$ Hamiltonian that is given in (\ref{5.2}).

We can now proceed as in the second-order scalar quantum field theory discussed above and represent the commutators by
\begin{eqnarray}
&&\left[z(\bar{k},t), -i\frac{\partial}{\partial z(\bar{k}^{\prime},t)}\right]=\delta^3(\bar{k}-\bar{k}^{\prime}),\qquad \left[x(\bar{k},t),-i\frac{\partial}{\partial x(\bar{k}^{\prime},t)}\right]=\delta^3(\bar{k}-\bar{k}^{\prime}).
\label{6.6}
\end{eqnarray}
With the vacuum obeying $a_1(\bar{k})\vert \Omega\rangle=0$, $a_2(\bar{k})\vert \Omega\rangle=0$ for each $\bar{k}$, from (\ref{6.2}) we obtain 
\begin{align}
&\langle z(\bar{k}),x(\bar{k})\vert a_1(\bar{k})\vert \Omega\rangle=\frac{1}{2(M_1^2-M_2^2)}\left[-\omega_2^2(\bar{k})z(\bar{k})+i\frac{\partial}{\partial x(\bar{k})}+i\omega_1(\bar{k})x(\bar{k})+\frac{1}{\omega_1(\bar{k})}\frac{\partial}{\partial z(\bar{k})}\right]\langle z(\bar{k}),x(\bar{k})\vert\Omega\rangle=0,
\nonumber\\
&\langle z(\bar{k}),x(\bar{k})\vert a_2(\bar{k})\vert \Omega\rangle=\frac{1}{2(M_1^2-M_2^2)}\left[\omega_1^2(\bar{k})z(\bar{k})-i\frac{\partial}{\partial x(\bar{k})}-i\omega_2(\bar{k})x(\bar{k})-\frac{1}{\omega_2(\bar{k})}\frac{\partial}{\partial z(\bar{k})}\right]\langle z(\bar{k}),x(\bar{k})\vert\Omega\rangle=0,
\label{6.7}
\end{align}
for each $\bar{k}$. From (\ref{6.7}) it follows that for each $\bar{k}$ we can identify each $\langle z(\bar{k}),x(\bar{k})\vert\Omega\rangle$ with the PU oscillator ground state wave function $\psi_0(z(\bar{k}),x(\bar{k}))$, which, analogously to (\ref{5.4}), is given by
\begin{align}
\psi_0(z(\bar{k}),x(\bar{k}))=\exp[\tfrac{1}{2}[\omega_1(\bar{k})+\omega_2(\bar{k})]\omega_1(\bar{k})\omega_2(\bar{k})z^2(\bar{k})+i\omega_1(\bar{k})\omega_2(\bar{k})z(\bar{k})x(\bar{k})-\tfrac{1}{2}[\omega_1(\bar{k})+\omega_2(\bar{k})]x^2(\bar{k})].
\label{6.8}
\end{align}
Consequently,  the normalization of the vacuum is given by
\begin{align}
\langle \Omega\vert \Omega\rangle&=\Pi_{\bar{k}}\int_{-\infty}^{\infty} dz(\bar{k})\int_{-\infty}^{\infty}dx(\bar{k})\langle \Omega\vert  z(\bar{k}),x(\bar{k})\rangle\langle z(\bar{k}),x(\bar{k})\vert\Omega\rangle
\nonumber\\
&=\Pi_{\bar{k}}\int_{-\infty}^{\infty} dz(\bar{k}) \int_{-\infty}^{\infty}dx(\bar{k}) \psi_0^*(z(\bar{k}),x(\bar{k}))\psi_0(z(\bar{k}),x(\bar{k})).
\label{6.9}
\end{align}
With each $\psi_0(z(\bar{k}),x(\bar{k}))$ diverging at large $z(\bar{k})$, we thus establish that the normalization of the field theory vacuum is infinite. Thus whatever is the normalization of the vacuum in the associated wave-mechanical limit translates into the same normalization in the quantum field theory.

\section{How to obtain a normalizable vacuum}
\label{S7}

\subsection{Similarity and symplectic transformations}
\label{S7A}

In analyzing the second-order plus fourth-order scalar field theory we note that with a conventional Hermitian field $\phi(x)$, and thus with $a_1^{\dagger}(\bar{k})$ and $a_2^{\dagger}(\bar{k})$ being the Hermitian conjugates of $a_1(\bar{k})$ and $a_2(\bar{k})$, the $a_2^{\dagger}(\bar{k})a_2(\bar{k})$ product would be positive definite  and  the energy spectrum of $H_S$ as given in (\ref{4.7}) would initially be unbounded from below, this being the familiar Ostrogradski instability of higher-derivative theories with Hermitian fields. However, from (\ref{4.6}) we see that $\langle \Omega\vert a_2(\bar{k})a_2^{\dagger}(\bar{k})\vert \Omega \rangle$  would be negative. This would  imply the potential presence of ghost  states of negative norm, with it then not being the case that a product such as $a_2(\bar{k})a_2^{\dagger}(\bar{k})$ could be positive definite. If one accepts this then  matrix elements of  the $-2(M_1^2-M_2^2)(\bar{k}^2+M_2^2)a^{\dagger}_2(\bar{k})a_2(\bar{k})$ term in $H_S$ would be compensated for by the ghost signature, and the energy spectrum of $H_S$ would then be bounded from below. While this takes care of the unboundedness from below of the energy spectrum, it appears to do so at a high price, namely the potential presence of unitarity-violating ghost states. But if $a_2^{\dagger}(\bar{k})$ is the Hermitian conjugate of $ a_2(\bar{k})$ then $\langle \Omega\vert a_2(\bar{k})a_2^{\dagger}(\bar{k})\vert \Omega \rangle$ would have to be positive. Thus despite the dagger notation $a_2^{\dagger}(\bar{k})$ could not be the Hermitian conjugate of $a_2(\bar{k})$. Hence our starting assumption that $\phi(x)$ is Hermitian could not be valid. Consequently, the Hamiltonian that is built out of the $\phi(x)$ field could not be Hermitian either. And in fact we have actually established that it is not, since the diverging of $\psi_0(z(\bar{k}),x(\bar{k}))$ at large $z(\bar{k})$ means that in an integration by parts we could not drop surface terms, with the presence of such surface terms preventing Hermiticity or self-adjointness.  With the eigenstates of the Hamiltonian not being normalizable, there not only are potential negative-norm states present, they are infinitely negative. 

Surprisingly, it is this very inability to drop surface terms in an integration by parts that actually saves the theory \cite{Bender2008a,Bender2008b}. Specifically, we have seen that we are working with a Hamiltonian $H_S$ (and likewise $H_{\rm PU}$) that is not Hermitian. However, all the energy eigenvalues associated with $H_S$ and $H_{\rm PU}$ are real.  Now common as its use is,  a Hermiticity condition is only sufficient to secure real eigenvalues but not necessary. (While Hermitian Hamiltonians have real eigenvalues, there is no converse theorem that says that a non-Hermitian Hamiltonian must have at least one  complex eigenvalue.) As to a  necessary condition, this has been found in \cite{bender2010PT,mannheim2018antilinearity}, with the necessary condition being that the Hamiltonian have an antilinear symmetry  \cite{footnote2}. The theory thus falls into the class of $PT$  theories ($P$ is the linear parity operator and $T$ is the antilinear time reversal operator) developed by Bender and collaborators \cite{,Bender1998,bender2007making,bender2019pt}. Critical to the $PT$ program  is that the wave functions be normalizable in some domain in the complex plane, a domain known technically as a Stokes wedge. Since the $\psi_0(z,x)$ and $\psi_0(z(\bar{k}), x(\bar{k}))$ wave functions are not normalizable with real $z$ or real $z(\bar{k})$, we have to take $z$ and $z(\bar{k})$ to be pure imaginary  in order to make the wave functions be normalizable \cite{footnote2a}. Then the theory is well-defined, with, as we discuss in Secs. \ref{S8} and \ref{S12}, the  Minkowski path integral accordingly also then being well behaved too \cite{Bender2008b,mannheim2018antilinearity}. 

However, rather than working with imaginary variables we can transform these pure imaginary variables into real ones by making a transformation  into the complex plane.   For the two-dimensional  $\widetilde{(z,p_z)}$ c-number column vector  this is achieved at the level of c-numbers by making the Poisson-bracket-preserving  symplectic transformation of the form (see e.g. \cite{mannheim2018antilinearity}) $S(\omega)=e^{-\omega \sigma_3}$ so that $z\rightarrow -iz\equiv y$, $p_z\rightarrow ip_z\equiv q$ when $\omega=i\pi/2$. (We have no need to modify $x$ or $x(\bar{k})$ since the wave functions already are well behaved when  $x$ or $x(\bar{k})$ become large.) With $z$ and consequently $p_z$ being pure imaginary it follows that the $y$ and $q$ 
c-numbers are real.

For the quantum operators we note that the transformation on the  $z$ and $p_z$ c-numbers converts the quantum-mechanical wave mechanics wave  operator associated with $H_{\rm PU}$  given in (\ref{5.2}) (and analogously $H_S$) from $H_{\rm PU}=-\partial_x^2/2 -ix\partial_z+(\omega_1^2+\omega_2^2)x^2/2-\omega_1^2\omega_2^2z^2/2$ into $\tilde{H}_{\rm PU}=-\partial_x^2/2 -x\partial_y+(\omega_1^2+\omega_2^2)x^2/2+\omega_1^2\omega_2^2y^2/2$. To determine a quantum operator form for $\tilde{H}_{\rm PU}$ we introduce the commutation-relation-preserving similarity transformations of the form
\begin{align}
S({\rm PU})=e^{\pi p_zz/2},\qquad S(S)=e^{\pi\int d^3x\pi_0(\bar{x},t)\phi(\bar{x},t)/2},
\label{7.1}
\end{align}
and obtain
\begin{align}
&S({\rm PU})zS({\rm PU})^{-1}=-iz\equiv y,\qquad S({\rm PU})p_zS({\rm PU})^{-1}=ip_z\equiv q,\qquad [y,q]=i,
\nonumber\\
& 
S(S)z(\bar{k})S(S)^{-1}=-iz(\bar{k})\equiv y(\bar{k}),\qquad S(S)p_z(\bar{k})S(S)^{-1}=ip_z(\bar{k})\equiv q(\bar{k}),
\qquad [y(\bar{k}),q(\bar{k}^{\prime})]=\delta^3(\bar{k}-\bar{k}^{\prime}).
\label{7.2}
\end{align}
In parallel with the classical symplectic transformation, for the quantum similarity transformation  we can set $S({\rm PU})=e^{-i\omega p_zz}$, with $\omega=i\pi/2$.
In terms  of these operators the quantum operator form for $\tilde{H}_{PU}$ is  given by $\tilde{H}_{\rm PU}=p_x^2/2 -iqx+(\omega_1^2+\omega_2^2)x^2/2+\omega_1^2\omega_2^2y^2/2$.
 With the $y$ and $q$  q-numbers being self-adjoint when acting on the eigenstates of  $\tilde{H}_{\rm PU}$,  and with the c-number $y$ and $q$ being real, following the symplectic and similarity transformations we see that, as had been noted in \cite{Bender2008a,Bender2008b},  the q-number $y$ and $q$ operators are Hermitian.  With this likewise holding for  the field theory case as well, the analog $y(\bar{k})$ and $q(\bar{k})$ will be Hermitian too, and thus now they are observable operators with real eigenvalues, just as one would want of a quantum field theory. Interestingly, despite now  being self-adjoint and having real eigenvalues, we see that neither of the transformed $\bar{H}_{\rm PU}$ or $\bar{H}_{\rm S}$ is Hermitian. However, as we  discuss below in Sec. \ref{S7D}, under a further similarity transformation we can bring them to  a Hermitian form \cite{Bender2008a,Bender2008b}.

In the analysis  presented in Sec. \ref{S5} the $H_{\rm PU}$ Hamiltonian  as given in (\ref{5.6}) has two realizations. In one $a_2$ annihilates the vacuum and in the other it is $a_2^{\dagger}$ that does so. These two choices correspond to totally different Hilbert spaces. Thus at the level of the quantum $H_{PU}$  there has to be some distinguishing feature between the two cases. In the $a_2^{\dagger}\vert \Omega \rangle=0$ case it is the initial $z$ and $p_z$ operators that are Hermitian, while in the $a_2\vert \Omega \rangle=0$ case it is the transformed $z$ and $p_z$ operators (viz. the $y$ and $q$ operators) that are Hermitian. At the level of the classical $H_{PU}$ we note that in the first case it is the initial $z$ and $p_z$ c-numbers that are real, while in the second case it is the transformed $z$ and $p_z$ c-numbers (viz. the $y$ and $q$ c-numbers) that are real. Below in Sec. \ref{S8} and \ref{S12B} we will see this same demarcation at the level of the path integral representation of the theory. All of these  remarks also apply to the $H_S$ field theory Hamiltonian. We now discuss the PU and field theory cases in detail.

\subsection{The PU case}
\label{S7B}

For the PU oscillator  (\ref{7.2})  leads to 
\begin{align}
\bar{H}_{\rm PU}&=\tfrac{1}{2}p_x^2(t)-iq(t)x(t)+\tfrac{1}{2}\left(\omega_1^2+\omega_2^2 \right)x^2(t)+\tfrac{1}{2}\omega_1^2\omega_2^2y^2(t),
\nonumber\\
 [y(t),q(t)]&=i, \qquad [x(t),p_x(t)]=i.
 \label{7.3}
 \end{align}
 Since the $\psi(y,x)$  wave functions are now well behaved at infinity, the $\bar{H}_{\rm PU}$ Hamiltonian is self-adjoint. However,  because of the $-iq(t)x(t)$ term it is not Hermitian. (To be Hermitian a Hamiltonian would have to obey $H_{ij}^*=H_{ji}$ in the basis in which it is self-adjoint.) Instead $\bar{H}_{\rm PU}$  is $PT$ symmetric. Specifically, with $p_x$ and $q$ being taken to be $PT$ even and $y$ and $x$ being taken to be $PT$ odd \cite{Bender2008b}, the $PT$ invariance of $\bar{H}_{\rm PU}$ and of the $[y,q]=i$ and $[x,p_x]=i$ commutators follows. (Under a similarity transformation not only do the operators transform,  so does $PT$ itself, doing so in such as way that whatever is the $PT$ behavior of the untransformed operator, then the similarity-transformed operator transforms the same way under the similarity-transformed $PT$ \cite{mannheim2018antilinearity}. While proving this result is trivial for linear operators, it is nontrivial for an antilinear operator such as $PT$. Thus like the original $PT$ odd coordinate operator $z$ the operator $y$ is $PT$ odd. Then from Hamilton's equations the $PT$ assignments of $x$, $p_x$ and $q$ follow.) Now when a Hamiltonian is not Hermitian the action of it to the right and the action of it to the left are not related by Hermitian conjugation. Thus in general one must distinguish between right- and left-eigenstates, both for the vacuum and the states that can be excited out of it. Thus we represent the $[y,q]=i$ and  $[x,p_x]=i$ commutators by  $q=-i\overrightarrow{\partial_y}$, $p_x=-i\overrightarrow{\partial_x}$ when acting to the right, and by $q=i\overleftarrow{\partial_y}$, $p_x=i\overleftarrow{\partial_x}$ when acting to the left. This then leads to  right- and left-ground-state wave functions of the bounded form \cite{Bender2008b}
\begin{align}
\psi_0^R(y,x)&=\exp[-\tfrac{1}{2}(\omega_1+\omega_2)\omega_1\omega_2y^2-\omega_1\omega_2yx-\tfrac{1}{2}(\omega_1+\omega_2)x^2]
\nonumber\\
&=\exp\left[-\frac{[(\omega_1+\omega_2)x+\omega_1\omega_2y]^2+\omega_1\omega_2(\omega_1^2+\omega_2^2+\omega_1\omega_2)y^2}{2(\omega_1+\omega_2)}\right],
\nonumber\\
\psi_0^L(y,x)&=\exp[-\tfrac{1}{2}(\omega_1+\omega_2)\omega_1\omega_2y^2+\omega_1\omega_2yx-\tfrac{1}{2}(\omega_1+\omega_2)x^2]
\nonumber\\
&=\exp\left[-\frac{[(\omega_1+\omega_2)x-\omega_1\omega_2y]^2+\omega_1\omega_2(\omega_1^2+\omega_2^2+\omega_1\omega_2)y^2}{2(\omega_1+\omega_2)}\right],
\label{7.4}
\end{align}
that converge for large positive or negative $y$ and $x$. 
Given these wave functions  the vacuum normalization is given by \cite{Bender2008b}
 \begin{align}
 \langle \Omega^{L}\vert \Omega^R\rangle&=\int_{-\infty}^{\infty} dy\int_{-\infty}^{\infty}dx\langle \Omega^{L}\vert y,x\rangle\langle y,x\vert\Omega^R\rangle=\int_{-\infty}^{\infty} dy\int_{-\infty}^{\infty}dx\psi_0^L(y,x)\psi_0^R(y,x)
 \nonumber\\
 &=\int_{-\infty}^{\infty} dy\int_{-\infty}^{\infty}dx\exp[-(\omega_1+\omega_2)\omega_1\omega_2y^2-(\omega_1+\omega_2)x^2]
 =\frac{\pi}{(\omega_1\omega_2)^{1/2}(\omega_1+\omega_2)},
 \label{7.5}
 \end{align}
 with the vacuum state thus being normalizable. In the following we shall understand the wave functions  to have been normalized to one, so that $\int dydx\psi_0^L(y,x)\psi_0^R(y,x)=1$ and  $\langle \Omega^{L}\vert \Omega^R\rangle=1$ \cite{footnote3}.

With the above $PT$ assignments and with $\dot{y}=i[\bar{H}_{\rm PU},y]=-ix$, $\dot{x}=p_x$, $\dot{p}_x=iq -(\omega_1^2+\omega_2^2)x$, $\dot{q}=-\omega_1^2\omega_2^2y$, we set
\begin{align}
y(t)&=-ia_1e^{-i\omega_1t}+a_2e^{-i\omega_2t}-i\hat{a}_1e^{i\omega_1t}+\hat{a}_2
e^{i\omega_2t},
\nonumber\\
x(t)&=-i\omega_1a_1e^{-i\omega_1t}+\omega_2a_2e^{-i\omega_2t}+i\omega_1\hat{a}_1
e^{i\omega_1t}-\omega_2\hat{a}_2e^{i\omega_2t},
\nonumber\\
p_x(t)&=-\omega_1^2a_1e^{-i\omega_1t}-i\omega_2^2a_2e^{-i\omega_2t}-\omega_1
^2\hat{a}_1e^{i\omega_1t}-i\omega_2^2\hat{a}_2e^{i\omega_2t},
\nonumber\\
q(t)&=\omega_1\omega_2[-\omega_2a_1e^{-i\omega_1t}-i\omega_1a_2e^{-i
\omega_2t}+\omega_2\hat{a}_1e^{i\omega_1t}+i\omega_1\hat{a}_2e^{i\omega_2t}],
\nonumber\\
a_1e^{-i\omega_1t}&=\frac{1}{2(\omega_1^2-\omega_2^2)}\left[-i\omega_2^2y(t)-p_x(t)+i\omega_1x(t)+\frac{q(t)}{\omega_1}\right],
\nonumber\\
\hat{a}_1e^{+i\omega_1t}&=\frac{1}{2(\omega_1^2-\omega_2^2)}\left[-i\omega_2^2y(t)-p_x(t)-i\omega_1x(t)-\frac{q(t)}{\omega_1}\right],
\nonumber\\
ia_2e^{-i\omega_2t}&=\frac{1}{2(\omega_1^2-\omega_2^2)}\left[i\omega_1^2y(t)+p_x(t)-i\omega_2x(t)-\frac{q(t)}{\omega_2}\right],
\nonumber\\
i\hat{a}_2e^{+i\omega_2t}&=\frac{1}{2(\omega_1^2-\omega_2^2)}\left[i\omega_1^2y(t)+p_x(t)+i\omega_2x(t)+\frac{q(t)}{\omega_2}\right].
\label{7.6}
\end{align}
In (\ref{7.6}) we have introduced $a_1$, $a_2$, $\hat{a}_1$ and $\hat{a}_2$, with the four creation and annihilation operators obeying $PTa_1TP=a_1$, $PTa_2TP=-a_2$, $PT\hat{a}_1TP=\hat{a}_1$, $PT\hat{a}_2TP=-\hat{a}_2$, so as to enforce the $PT$ assignments of $y$, $x$, $p_x$ and $q$. Comparing with (\ref{5.5}) we have $(a_1,a_2,a_1^{\dagger},a^{\dagger}_2)\rightarrow (a_1,ia_2,\hat{a}_1,i\hat{a}_2)$.

With (\ref{7.3}) and (\ref{7.6}) the Hamiltonian is given by 
\begin{equation}
\bar{H}_{\rm PU}=2(\omega_1^2-\omega_2^2)\left(\omega_1^2\hat{a}_1a_1+\omega_2^2
\hat{a}_2a_2\right)+\tfrac{1}{2}(\omega_1+\omega_2),
\label{7.7}
\end{equation}
and with $[y(t),q(t)]=i$, $[x(t),p_x(t)]=i$, a faithful representation of the operator commutation algebra is given by
\begin{align}
&[a_1,\hat{a}_1]=\frac{1}{2\omega_1(\omega_1^2-\omega_2^2)},\quad
[a_2,\hat{a}_2]=\frac{1}{2\omega_2(\omega_1^2-\omega_2^2)},
\nonumber\\
&[a_1,a_2]=0,\quad[a_1,\hat{a}_2]=0,\quad[\hat{a}_1,a_2]=0,\quad
[\hat{a}_1,\hat{a}_2]=0.
\label{7.8}
\end{align}
In comparing (\ref{7.8}) with (\ref{5.7}) we see that the sign of the $[a_2,\hat{a}_2]$ commutator  has changed (so that it is now positive), while the sign of the $[a_1,\hat{a}_1]$ commutator has not (so that it remains positive). It is because we only continued one of the two oscillators into the complex plane (the one associated with $z$ and $p_z$  and not the the one associated with $x$ and $p_x$) that  the sign of just one of the $[a_1,\hat{a}_1]$ and $[a_2,\hat{a}_2]$  commutators is changed. As we see, the continuation into the complex plane resolves the negative norm problem.

With the $PT$ assignments of $a_1$, $a_2$, $\hat{a}_1$ and $\hat{a}_2$, we confirm the $PT$ invariance of (\ref{7.7}) and (\ref{7.8}).
In (\ref{7.7}) and (\ref{7.8}) the relative signs are all positive (we take $\omega_1>\omega_2>0$ for definitiveness),  so these
equations define a standard positive energy, positive norm, two-dimensional harmonic oscillator system. Given the creation and annihilation operators the left- and right-vacua are defined by
\begin{align}
\langle \Omega^{L}\vert \hat{a}_1=0, \qquad \langle \Omega^{L}\vert \hat{a}_2=0,\qquad a_1\vert \Omega^R\rangle=0,\qquad a_2\vert \Omega^R\rangle=0.
\label{7.9}
\end{align}
By exciting modes out of the left- and right-vacua we can build excited states that have positive norm \cite{Bender2008a}, viz. $\langle n^L\vert m^R\rangle=\delta_{nm}$, and obey a completeness relation 
\begin{eqnarray}
\sum \vert n_1^R\rangle\langle n_1^{L}\vert +\sum \vert n_2^R\rangle\langle n_2^{L}\vert =I.
\label{7.10}
\end{eqnarray}
Even though these norms are all positive, due to the presence of the factor $i$ in the $(a_1,\hat{a}_1)$ sector of $y(t)$ as given in  (\ref{7.6}), the insertion of (\ref{7.10}) into $-i\langle\Omega^{L}\vert T[y(t)y(0)]\vert \Omega^R \rangle$ (corresponding to  $+i\langle\Omega^{L}\vert T[z(t)z(0)]\vert \Omega^R \rangle$)  generates \cite{Bender2008b} the relative minus sign in  the nonrelativistic limit of the $-[1/(k^2-M_1^2)-1/(k^2-M_2^2)]/(M_1^2-M_2^2)$ propagator given in (\ref{4.2}), viz. $-[1/(\omega^2-\omega_1^2)-1/(\omega^2-\omega_2^2)]/(\omega_1^2-\omega_2^2)$. We thus establish the consistency and physical viability of the similarity-transformed $PU$ oscillator theory. 

In this context it is of interest to recall  a study by Pauli \cite{Pauli1943}.  In this study Pauli transformed the position and momentum operators into the complex plane. His interest was in converting a positive definite quantum metric theory into an indefinite metric one. The work of \cite{Bender2008a,Bender2008b} described here does the same thing but in reverse, converting an indefinite metric quantum theory into a positive definite metric one.

\subsection{The relativistic case}
\label{S7C}

For $\bar{H}_S$ we introduce  creation and annihilation operators for $\bar{\phi}=S(S)\phi S(S)^{-1}=-i\phi(x)$  of the form
\begin{align}
\bar{\phi}(x)=\int \frac{d^3k}{(2\pi)^{3/2}}\left [-ia_1(\bar{k})e^{-i\omega_1(\bar{k})t+i\bar{k}\cdot \bar{x}}+a_2(\bar{k})e^{-i\omega_2(\bar{k}) t+i\bar{k}\cdot \bar{x}}-i\hat{a}_1(\bar{k})e^{i\omega_1(\bar{k}) t-i\bar{k}\cdot \bar{x}}+\hat{a}_2(\bar{k})e^{i\omega_2(\bar{k}) t-i\bar{k}\cdot \bar{x}}\right].
\label{7.11}
\end{align}
Comparing with (\ref{4.5}) we have $(a_1(\bar{k}),a_2(\bar{k}),a_1^{\dagger}(\bar{k}),a^{\dagger}_2(\bar{k}))\rightarrow (a_1(\bar{k}),ia_2(\bar{k}),\hat{a}_1(\bar{k}),i\hat{a}_2(\bar{k}))$. Since scalar fields are $PT$ even while coordinates are $PT$ odd, then unlike the $PT$ odd $y(t)$, $\bar{\phi}(\bar{x},t)$ is $PT$ even, i.e., $PT\bar{\phi}(\bar{x},t)TP=\bar{\phi}(-\bar{x},-t)$. From (\ref{7.11}) it follows that $PTa_1(\bar{k})TP=-a_1(\bar{k})$, $PTa_2(\bar{k})TP=a_2(\bar{k})$, $PT\hat{a}_1(\bar{k})TP=-\hat{a}_1(\bar{k})$, $PT\hat{a}_2(\bar{k})TP=\hat{a}_2(\bar{k})$. Given (\ref{7.11}), the field-theoretic Hamiltonian and a faithful representation of the commutation relations  are given by \cite{Bender2008b}
\begin{eqnarray}
\bar{H}_S&=&\frac{1}{2}\int d^3k\bigg{[}2(M_1^2-M_2^2)(\bar{k}^2+M_1^2)
\left[\hat{a}_{1}(\bar{k})a_1(\bar{k})+a_{1}(\bar{k})\hat{a}_1(\bar{k})\right]
\nonumber\\
&+&2(M_1^2-M_2^2)(\bar{k}^2+M_2^2)\left[\hat{a}_{2}(\bar{k})a_2(\bar{k})+a_{2}(\bar{k})\hat{a}_2(\bar{k})\right]\bigg{]},
\label{7.12}
\end{eqnarray}
and
\begin{eqnarray}
&& [\dot{\bar{\phi}}(\bar{x},t),\bar{\phi}(0)]=0,\qquad [\ddot{\bar{\phi}}(\bar{x},t),\bar{\phi}(0)]=0,\qquad [\dddot{\bar{\phi}}(\bar{x},t),\bar{\phi}(0)]=i\delta^3(\bar{x}),
\nonumber\\
&&[a_1(\bar{k}),\hat{a}_{1}(\bar{k}^{\prime})]=[2(M_1^2-M_2^2)(\bar{k}^2+
M_1^2)^{1/2}]^{-1}\delta^3(\bar{k}-\bar{k}^{\prime}),
\nonumber\\
&&[a_2(\bar{k}),\hat{a}_{2}(\bar{k}^{\prime})]=[2(M_1^2-M_2^2)(\bar{k}^2+
M_2^2)^{1/2}]^{-1}\delta^3(\bar{k}-\bar{k}^{\prime}),
\nonumber\\
&&[a_1(\bar{k}),a_{2}(\bar{k}^{\prime})]=0,
\quad [a_1(\bar{k}),\hat{a}_{2}(\bar{k}^{\prime})]=0,
\quad [\hat{a}_{1}(\bar{k}),a_{2}(\bar{k}^{\prime})]=0,
\quad [\hat{a}_{1}(\bar{k}),\hat{a}_{2}(\bar{k}^{\prime})]=0.
\label{7.13}
\end{eqnarray}
With the $PT$ assignments of the creation and annihilation operators we check that $\bar{H}_S$ is $PT$ even, while the commutation relations respect $PT$ symmetry. Thus as constructed, the Hamiltonian is $PT$ even. (In general, even without Hermiticity one still has $CPT$ symmetry \cite{mannheim2018antilinearity}, but since the fields are neutral $C$ is separately conserved, so in this case $CPT$ defaults to $PT$.) 

With all relative signs in (\ref{7.12}) and (\ref{7.13}) being positive (we take $M_1^2>M_2^2$ for definitiveness), there are no states of negative norm or of negative energy. The discussion completely parallels that of the $PU$ oscillator model  given above. We introduce
\begin{align}
y(\bar{k},t)&=-ia_1(\bar{k})e^{-i\omega_1(\bar{k})t}+a_2(\bar{k})e^{-i\omega_2(\bar{k})t}-i\hat{a}_1(\bar{k})e^{i\omega_1(\bar{k})t}+\hat{a}_2(\bar{k})
e^{i\omega_2(\bar{k})t},
\nonumber\\
x(\bar{k},t)&=-i\omega_1(\bar{k})a_1(\bar{k})e^{-i\omega_1(\bar{k})t}+\omega_2(\bar{k})a_2(\bar{k})e^{-i\omega_2(\bar{k})t}+i\omega_1(\bar{k})\hat{a}_1(\bar{k})
e^{i\omega_1(\bar{k})t}-\omega_2(\bar{k})\hat{a}_2(\bar{k})e^{i\omega_2(\bar{k})t},
\nonumber\\
p_x(\bar{k},t)&=-\omega_1^2(\bar{k})a_1(\bar{k})e^{-i\omega_1(\bar{k})t}-i\omega_2^2(\bar{k})a_2(\bar{k})e^{-i\omega_2(\bar{k})t}-\omega_1
^2(\bar{k})\hat{a}_1(\bar{k})e^{i\omega_1(\bar{k})t}-i\omega_2^2(\bar{k})\hat{a}_2(\bar{k})e^{i\omega_2(\bar{k})t},
\nonumber\\
q(\bar{k},t)&=\omega_1(\bar{k})\omega_2(\bar{k})[-\omega_2(\bar{k})a_1(\bar{k})e^{-i\omega_1(\bar{k})t}-i\omega_1(\bar{k})a_2(\bar{k})e^{-i\omega_2(\bar{k})t}+\omega_2(\bar{k})\hat{a}_1(\bar{k})e^{i\omega_1(\bar{k})t}+i\omega_1(\bar{k})\hat{a}_2(\bar{k})e^{i\omega_2(\bar{k})t}],
\label{7.14}
\end{align}
with the $PT$-symmetric $\bar{H}_S$ then taking the form
\begin{align}
\bar{H}_S=\int d^3k\bigg{[}\frac{p_x^2(\bar{k},t)}{2}-iq(\bar{k},t)x(\bar{k},t)+\frac{1}{2}\left[\omega_1^2(\bar{k})+\omega_2^2(\bar{k}) \right]x^2(\bar{k},t)+\frac{1}{2}\omega_1^2(\bar{k})\omega_2^2(\bar{k})y^2(\bar{k},t)\bigg{]}.
\label{7.15}
\end{align}
With the $PT$ assignments of the creation and annihilation operators $y(\bar{k},t)$ and $x(\bar{k},t)$ are $PT$ even, while $p_x(\bar{k},t)$ and $q(\bar{k},t)$ are $PT$ odd.
In analog to (\ref{7.4}) the left- and right-ground-state wave functions are given by 
\begin{align}
\psi_0^R(y(\bar{k}),x(\bar{k}))&=\exp[-\tfrac{1}{2}(\omega_1(\bar{k})+\omega_2(\bar{k}))\omega_1(\bar{k})\omega_2(\bar{k})y^2(\bar{k})-\omega_1(\bar{k})\omega_2(\bar{k})y(\bar{k})x(\bar{k})-\tfrac{1}{2}(\omega_1(\bar{k})+\omega_2(\bar{k}))x^2(\bar{k})],
\nonumber\\
\psi_0^L(y(\bar{k}),x(\bar{k}))&=\exp[-\tfrac{1}{2}(\omega_1(\bar{k})+\omega_2(\bar{k}))\omega_1(\bar{k})\omega_2(\bar{k})y^2(\bar{k})+\omega_1(\bar{k})\omega_2(\bar{k})y(\bar{k})x(\bar{k})-\tfrac{1}{2}(\omega_1(\bar{k})+\omega_2(\bar{k}))x^2(\bar{k})].
\label{7.16}
\end{align}
Introducing left- and right-vacua that obey
\begin{align}
\langle \Omega^{L}\vert \hat{a}_1(\bar{k})=0, \qquad \langle \Omega^{L}\vert \hat{a}_2(\bar{k})=0,\qquad a_1(\bar{k})\vert \Omega^R\rangle=0,\qquad a_2(\bar{k})\vert \Omega^R\rangle=0
\label{7.17}
\end{align}
for all $\bar{k}$, we find that 
\begin{align}
\langle \Omega^{L}\vert \bar{H}_S\vert\Omega^R\rangle&=\int d^3k\bigg{[}\frac{1}{2}(\bar{k}^2+M_1^2)^{1/2}+\frac{1}{2}(\bar{k}^2+M_2^2)^{1/2}\bigg{]}\delta^3(\bar{0}),
\nonumber\\
\langle \Omega^{L}\vert \Omega^R\rangle&=\Pi_{\bar{k}}\int_{-\infty}^{\infty} dy(\bar{k})\int_{-\infty}^{\infty}dx(\bar{k})\langle \Omega^{L}\vert  y(\bar{k}),x(\bar{k})\rangle\langle y(\bar{k}),x(\bar{k})\vert\Omega^{R}\rangle
\nonumber\\
&=\Pi_{\bar{k}}\int_{-\infty}^{\infty} dy(\bar{k}) \int_{-\infty}^{\infty}dx(\bar{k}) \psi_0^L(y(\bar{k}),x(\bar{k}))\psi_0^R(y(\bar{k}),x(\bar{k}))=
\Pi_{\bar{k}}1=1.
\label{7.18}
\end{align}
We thus confirm that the vacuum normalization is both finite and positive, while the vacuum energy has the conventional zero-point infinity associated with an infinite number of modes. (This infinity occurs because $\bar{H}_S$ contains an infinite number of modes  and not because $\langle \Omega^{L}\vert \Omega^R\rangle$ itself is infinite.) We thus establish the consistency and physical viability of the similarity-transformed higher-derivative scalar field theory. And we note that even though all the norms are  positive, due to the presence of the factor $i$ in the $(a_1,\hat{a}_1)$ sector of $y(\bar{k},t)$ as given in  (\ref{7.14}), the insertion of (\ref{7.10}) into $-i\langle\Omega^{L}\vert T[\bar{\phi}(x)\bar{\phi}(0)]\vert \Omega^R \rangle$ (corresponding to  $+i\langle\Omega^{L}\vert T[\phi(x)\phi(0)]\vert \Omega^R \rangle$) generates the relative minus sign in $-[1/(k^2-M_1^2)-1/(k^2-M_2^2)]/(M_1^2-M_2^2)$ \cite{Bender2008b}. Thus with one similarity transform into an appropriate Stokes wedge we solve both the vacuum normalization problem and the negative-norm problem. 

At this point we can see the key aspect of our study. Ordinarily in quantum field theory it is taken as a given that one should use the Dirac inner product $\langle \Omega\vert \Omega\rangle$, viz. $\langle \Omega^R\vert \Omega^R\rangle$, for the vacuum. And also it is taken as a given that this inner product is finite. In this paper we have provided a procedure for checking whether this is in fact the case, and presented a second-order-derivative  plus fourth-order-derivative  model  in which it explicitly is not finite. For this particular model we have found a different inner product, viz. $\langle \Omega^{L}\vert \Omega^R\rangle$, that is finite. (For a Hamiltonian that is Hermitian $\vert \Omega^R\rangle=\vert \Omega\rangle$, $\langle \Omega^L\vert =\langle \Omega\vert$, and $\langle \Omega^L\vert \Omega^R\rangle=\langle \Omega\vert \Omega\rangle$.)  
And thus in general one has to determine whether or not $\langle \Omega^R\vert \Omega^R\rangle$ is finite on  case by case basis.  

We should note that in general there is no requirement that a quantum field theory must use the Dirac  inner product $\langle \Omega \vert \Omega \rangle$. What one does need of an acceptable inner product is that it be finite, positive and time independent. And as long as we can find one we have a well-defined Hilbert space. However, if the Hamiltonian is not Hermitian then the Dirac inner product obeys $\langle \Omega(t) \vert \Omega(t) \rangle =  \langle \Omega(t=0) \vert e^{iH^{\dagger}t}e^{-iHt}\vert \Omega(t=0) \rangle \neq  \langle \Omega(t=0) \vert \Omega(t=0) \rangle$, to thus not be time independent, and thus not acceptable. 

In regard to time independence, we also note  that  in the nonrelativistic case the time derivative of the $\int d^3x \psi^*(x)\psi(x)$ probability is related to an asymptotic spatial surface term. The  time independence of $\int d^3x \psi^*(x)\psi(x)$ is then secured if the asymptotic spatial surface term vanishes, i.e., the time independence of $\int d^3x \psi^*(x)\psi(x)$ is determined by asymptotic spatial boundary conditions.  If it turns out that the theory is well behaved asymptotically for Hermitian  fields then one can use the Dirac norm. If it is not well behaved asymptotically for Hermitian fields, then the insight of the $PT$ program is to look to see if instead there then might be some domain in the complex plane where the theory is well behaved asymptotically. If there is, then one can have a good Hilbert space, but the inner product will not then be the Dirac norm.  If the theory has a $PT$ symmetry, then the norm will be the $PT$ theory norm, and when the Hamiltonian is not Hermitian the left-right inner product obeys $\langle \Omega^L(t) \vert \Omega^R(t) \rangle =  \langle \Omega^L(t=0) \vert e^{iHt}e^{-iHt}\vert \Omega^R(t=0) \rangle=  \langle \Omega^L(t=0) \vert \Omega^R(t=0) \rangle$
to thus  be time independent. And this is the case for the  second-order plus fourth-order theory. 

From our treatment of $H_{\rm PU}$ given in (\ref{5.2}) and of $H_S$ given in (\ref{6.5}) we see the secret of $PT$ theory. As constructed, both of these Hamiltonians look to be Hermitian, as they are composed of operators each one of which is separately Hermitian. However, while these operators are self-adjoint when acting on their own eigenstates, they are not self-adjoint when acting on the eigenstates of $H_{\rm PU}$ or $H_S$. It is only after a continuation into the complex plane that one can find a domain in which $\bar{H}_{\rm PU}$ or $\bar{H}_S$ and the operators in them (such as $\bar{\phi}(x)$) are all self-adjoint when acting on the eigenstates of $\bar{H}_{\rm PU}$ or $\bar{H}_S$. In contrast, in a standard Hermitian theory  the Hamiltonian and the operators in it are all self-adjoint when acting on the eigenstates of the Hamiltonian, and no continuation into the complex plane is needed.

Once we have established that the Hilbert space for  the free theory $\bar{H}_S$ has a left-right inner product that is both finite and positive, the inner product must remain finite and positive when we include renormalizable interactions, since one cannot change the signature of a Hilbert space  or generate uncancellable infinites in any perturbative order. However, given the relative minus sign in $-[1/(k^2-M_1^2)-1/(k^2-M_2^2)]/(M_1^2-M_2^2)$, loop diagrams will contain contributions with negative discontinuities, and as such they would on their own violate unitarity. Now as we immediately show in Sec. \ref{S7D}, for a right-eigenstate $\vert n\rangle$ of  $\bar{H}_S$ the left-eigenstate is  given not by $\langle n \vert$ but by $\langle n\vert e^{-Q}$, where $Q$ is given in (\ref{7.19}) (as generalized to $y(\bar{k})$, $x(\bar{k})$, $q(\bar{k})$, $p_z(\bar{k})$). The operator $Q$ is itself modified in perturbation theory,  and it is this modification to the tree approximation graph that enables the theory to cancel the negative loop discontinuities \cite{Mannheim2018}, so that all resulting discontinuities are unitarity-preserving positive ones. In this way then unitarity is preserved in the presence of interactions.
 
With the finiteness or otherwise of the path integral also being determined by spatial boundary conditions, in Sec. \ref {S8} we  discuss our findings from the perspective of path integrals. In Sec. \ref{S12} we augment our path integral results with a quick and  straightforward  analysis  of  Feynman's $i\epsilon$ prescription for propagators. However before doing so we show how to write 
$\bar{H}_{PU}$ and analogously $\bar{H}_S$ in a manifestly Hermitian form.

\subsection{Transforming to a Hermitian Hamiltonian}
\label{S7D}

While not obeying $H^*_{ij}=H_{ji}$ in the basis of its eigenfunctions, the $\bar{H}_{PU}$ Hamiltonian given in (\ref{7.3}) is self-adjoint and has all eigenvalues real. In addition its eigenspectrum is complete (a complete set of polynomial functions of $x$ and $y$ times the ground state wave functions given in (\ref{7.4}) \cite{Bender2008a}). Thus by a similarity transformation $\bar{H}_{PU}$  can be brought to a basis in which $H^*_{ij}=H_{ji}$. (In general if $H^{\prime}=H^{\prime \dagger}$ then under a similarity but not unitary transformation of the form $H^{\prime}=SHS^{-1}$ we have $SHS^{-1}=S^{-1 \dagger}H^{\dagger}S^{\dagger}$, i.e, $H^{\dagger}=S^{\dagger}SH(S^{\dagger}S)^{-1}$, with the relation $H^{\prime}=H^{\prime \dagger}$ not being invariant under a non-unitary similarity transformation.) 
As shown in \cite{Bender2008a,Bender2008b}, for $H_{\rm PU}$ we introduce  
\begin{eqnarray}
Q=\alpha p_xq+\alpha\omega_1^2\omega_2^2 xy,\qquad \alpha=\frac{1}{\omega_1\omega_2}{\rm log}\left(\frac{\omega_1+\omega_2}{\omega_1-\omega_2}\right),
\label{7.19}
\end{eqnarray}
with the requisite transformation then being  given  by
\begin{align}
e^{-Q/2}ye^{Q/2}&=y\cosh\theta+i(\omega_1\omega_2)^{-1}p_x\sinh\theta,
\nonumber\\
e^{-Q/2}xe^{Q/2}&=x\cosh\theta+i(\omega_1\omega_2)^{-1}q\sinh\theta,
\nonumber\\
e^{-Q/2}pe^{Q/2}&=p_x\cosh\theta-i(\omega_1\omega_2)y\sinh\theta,
\nonumber\\
e^{-Q/2}qe^{Q/2}&=q\cosh\theta-i(\omega_1\omega_2)x\sinh\theta,
\nonumber\\
e^{-Q/2}\bar{H}_{PU}e^{Q/2}&=\bar{H}^{\prime}_{PU}
=\frac{p_x^2}{2}+\frac{q^2}{2\omega_1^2}+
\frac{1}{2}\omega_1^2x^2+\frac{1}{2}\omega_1^2\omega_2^2y^2,
\label{7.20}
\end{align}
where $\theta=\alpha\omega_1\omega_2/2={\rm arc\sinh}[\omega_2/(\omega_1^2-\omega_2^2)^{1/2}]$.
We recognize $\bar{H}_{\rm PU}^{\prime}$ as being a fully acceptable standard, positive norm  two-dimensional oscillator system, one for which we can use the Dirac inner product. Moreover, by the analysis given for the harmonic oscillator in Sec. \ref{S2}, it follows that in this basis the Dirac norm of the vacuum is finite. Similarly, by extension, following an analogous transformation  for each $\bar{k}$, so is the vacuum Dirac norm of the second-order-derivative plus fourth-order derivative scalar field theory.

In addition we note that with its phase being $-Q/2$ rather than $-iQ/2$, the $e^{-Q/2}$ operator is not unitary. The transformation from $\bar{H}_{PU}$ to $\bar{H}^{\prime}_{PU}$ is thus not a unitary transformation, but is a transformation from a skew basis with eigenvectors $|n\rangle$ to an orthogonal basis with eigenvectors 
\begin{align}
|n^{\prime}\rangle=e^{-Q/2}|n\rangle,~~~~\langle n^{\prime}|=\langle n|e^{-Q/2}.
\label{7.21	}
\end{align}
Then since $\langle n^{\prime}|m^{\prime}\rangle =\delta_{mn}$, the eigenstates of $\bar{H}$ obey
\begin{eqnarray}
&&\langle n|e^{-Q}|m\rangle=\delta_{mn},\quad \sum_n|n\rangle\langle n|e^{-Q}=I,
\nonumber\\
&& \bar{H}=\sum _n|n\rangle E_n\langle n|e^{-Q}, \quad
\bar{H}|n\rangle=E_n|n\rangle,\quad \langle n|e^{-Q}\bar{H}=\langle n|e^{-Q}E_n,
\label{7.22}
\end{eqnarray}
with $\vert n\rangle$ being a right-eigenstate of $\bar{H}$ and $\langle n\vert  e^{-Q}$ being left-eigenstate of $\bar{H}$.
We thus recognize the inner product as being not $\langle n|m\rangle$ but $\langle n|e^{-Q}|m\rangle\equiv \langle L\vert R\rangle$, with the required conjugate of $|n\rangle$ being $\langle n|e^{-Q}$. This state is also the $PT$ conjugate of $|n\rangle$, so that the inner product is the overlap of a state with its $PT$ conjugate rather than that with its Hermitian conjugate, just as we had noted earlier. And as such this inner product is positive definite since $\langle n^{\prime}|m^{\prime}\rangle =\delta_{mn}$ is. The PU oscillator theory (and by analog the scalar quantum field theory) is thus a fully viable unitary theory. Thus starting from $H_{PU}$ given in (\ref{5.2}) we only need make two similarity transformations, viz.  (\ref{7.2}) and (\ref{7.20}), in order to be able to establish that the theory is free of negative norm states, and has a vacuum with a finite and positive norm. From the form given in (\ref{7.20}) for $\bar{H}^{\prime}_{PU}$ it follows that all the operators in it are observable quantum operators, with all experimental measurements then only involving the real quantities that are their eigenvalues.

\subsection{The invisible factor of $i$ and the reality of the classical limit}
\label{S7E}

As we have seen, the factor of $i$ that is present in $\bar{H}_{\rm PU}$ as given in (\ref{7.3}) is not present in $\bar{H}_{\rm PU}^{\prime}$ as given in (\ref{7.20}). It should thus be possible to show that its presence is not relevant even without making the transformation  given (\ref{7.20}). To this end we determine the Hamilton equations of motion associated with $\bar{H}_{\rm PU}$, and find them to be of the form 
\begin{align}
\dot{y}=i[\bar{H}_{\rm PU},y]=-ix, \quad \dot{x}=p_x, \quad \dot{p}_x=iq-(\omega_1^2+\omega_2^2)x, \quad \dot{q}=-\omega_1^2\omega_2^2y.
\label{7.23}
\end{align}
From these relations we obtain equations of motion for each dynamical variable of the form
\begin{align}
\ddddot{y}+(\omega_1^2+\omega_2^2)\ddot{y}+\omega_1^2\omega_2^2y=0,\qquad \ddddot{x}+(\omega_1^2+\omega_2^2)\ddot{x}+\omega_1^2\omega_2^2x=0,
\nonumber\\
\ddddot{p_x}+(\omega_1^2+\omega_2^2)\ddot{p}_x+\omega_1^2\omega_2^2p_x=0,\qquad \ddddot{q}+(\omega_1^2+\omega_2^2)\ddot{q}+\omega_1^2\omega_2^2q=0.
\label{7.24}
\end{align}
Thus not  only do all four dynamical variables obey the same fourth-order-derivative equation of motion (and not only that, they obey the same equation of motion as the original one given in (\ref{5.1})), as we see, no factor $i$ appears in any of them. Thus all the factors of $i$ that appear in (\ref{7.23}) are removed in (\ref{7.24}).

To reinforce this result we determine the $\bar{L}_{\rm PU}$ Lagrangian as the Legendre transform of $\bar{H}_{\rm PU}$, viz.
\begin{align}  
\bar{ L}_{\rm PU}=p_x\dot{x}+q\dot{y}-\tfrac{1}{2}p_x^2+iqx-\tfrac{1}{2}\left(\omega_1^2+\omega_2^2 \right)x^2-\tfrac{1}{2}\omega_1^2\omega_2^2y^2.
 \label{7.25}
 \end{align}
Unconstrained variation of the action $\bar{I}_{\rm PU}=\int dt \bar{L}_{\rm PU}$ with respect to $p_x$ and $q$ and variation with respect to $y$ and $x$ with their values held fixed at the endpoints of $\bar{I}_{\rm PU}$ then leads us  right back to (\ref{7.23}), just as it should. 
Using Hamilton's equations of motion to eliminate $p_x$ and $q$ we obtain 
\begin{align}  
\bar{ L}_{\rm PU}=\tfrac{1}{2}\dot{x}^2-\tfrac{1}{2}\left(\omega_1^2+\omega_2^2 \right)x^2-\tfrac{1}{2}\omega_1^2\omega_2^2y^2,
 \label{7.26}
 \end{align}
and thus obtain  a Lagrangian that contains no factors of $i$ at all. 

Now we note that if we additionally even use the $\dot{y}=-ix$ equation as well in order to eliminate $x$ we obtain
\begin{align}  
\bar{ L}_{\rm PU}=-\tfrac{1}{2}\ddot{y}^2+\tfrac{1}{2}\left(\omega_1^2+\omega_2^2 \right)\dot{y}^2-\tfrac{1}{2}\omega_1^2\omega_2^2y^2,
 \label{7.27}
 \end{align}
with Euler-Lagrange variation then recovering $\ddddot{y}+(\omega_1^2+\omega_2^2)\ddot{y}+\omega_1^2\omega_2^2y=0$. While this procedure does gives the correct equation of motion for $y$, it only involves a restricted variation in which $x$ is set equal to $i\dot{y}$ on every variational path. Nonetheless, if our only objective is to obtain the classical equations of motion, we could in fact start with  (\ref{7.27}). Moreover,  we could not start with (\ref{7.26}) and do an unconstrained variation on $x$ and $y$, as we would not get the correct Hamilton equations of motion.  (We would instead get $y=0$ and $\ddot{x}+(\omega_1^2+\omega_2^2)x=0$.) Thus in order to get the correct equations of motion we must start not with (\ref{7.26}) but with the phase space (\ref{7.25}) that depends on all four of the dynamical variables $y$, $x$, $q$ and $p_x$. Since we originally introduced a Lagrange multiplier for $\dot{z}-x$ and then used the method of Dirac constraints \cite{Mannheim2000,Mannheim2005} in order to obtain the phase-space-based $H_{\rm PU}$ Hamiltonian given in (\ref{5.2}) in the first place, the PU oscillator theory is intrinsically a Hamiltonian-based theory rather than a Lagrangian-based one

However, for the quantum theory starting with (\ref{7.27}) and doing a constrained variation  is far too restrictive as it leaves out the whole family of $x \neq i\dot{y}$ paths that are needed for the path integral quantization procedure that is described below in Secs. \ref{S8} and \ref{S12}. With the theory being Hamiltonian-based  the path integral measure must be based on integrating over all $y$, $x$, $q$ and $p_x$ paths. However, the path integration over $p_x$ and $q$ is straightforward and leaves us with a path integral with an action based on (\ref{7.26}). And with this (\ref{7.26}) we must then integrate over all $x$ and $y$ paths independently. As we  show in Sec. \ref{S8}, this same  $z\rightarrow iy$ transformation but with $x$ not transformed is needed in order to make the path integral exist at both the PU and scalar field level. In both the PU and scalar field cases we finish up with a path integral whose requisite classical action is based on a c-number $\bar{ L}_{\rm PU}$ or its scalar field analog that is built out of real c-number fields and real coefficients. The factor of $i$ that appears in  $\bar{H}_{\rm  PU}$ or $\bar{H}_{S}$ at either the classical or quantum level thus does not appear in the path integral at all, and thus not in the Green's functions that are obtained from it.

A somewhat curious feature of (\ref{7.27}) is that with real $y$ $\bar{L}_{\rm PU}$ has the opposite overall sign to that of the original Lagrangian given in (\ref{5.1}) as evaluated with real $z$, viz.
\begin{align}
L_{\rm PU}=\tfrac{1}{2}{\ddot z}^2-\tfrac{1}{2}\left(\omega_1^2
+\omega_2^2\right){\dot z}^2+\tfrac{1}{2}\omega_1^2\omega_2^2z^2.
 \label{7.28}
 \end{align}
This would initially suggest that the energy eigenvalues might change sign in the $z\rightarrow iy$ transformation. That this could not in principle be the case is because the equations of motion are not sensitive to the overall sign of the action, and thus the  frequencies that obey the equations of motion are not changed. That this not actually is the case will be explained in Sec. \ref{S12D}, where it is shown that there is a compensating change in sign in the canonical commutators.

We would like to note that while we have used $L_{\rm PU}$ and the $H_{\rm PU}$ in order to find  $\bar{L}_{\rm PU}$ and the self-adjoint $\bar{H}_{\rm PU}$, once we have $\bar{H}_{\rm PU}$ and $\bar{L}_{\rm PU}$ we can use them ab initio to define the theory and have no need to refer to $H_{\rm PU}$ and $L_{\rm PU}$ at all. All the operators that appear in the quantum $\bar{H}_{\rm PU}$ including $\bar{H}_{\rm PU}$ itself are self-adjoint and have real eigenvalues, with the eigenvalues of the quantum $y$ and $x$ being the classical fields that appear in the classical $\bar{L}_{\rm PU}$. Consequently the classical limit is real. Moreover, 
 if we do start ab initio  with  $\bar{H}_{\rm PU}$ and $\bar{L}_{\rm PU}$, we can then define interactions in terms of the operators that appear in them. Thus in the application of our approach to second-order-derivative plus fourth-order-derivative gravity that we present  in Sec. \ref{S11}  it is $y\equiv \bar{g}_{\mu\nu}$ and not $z\equiv g_{\mu\nu}$ that will represent the metric. Moreover, for determining the classical equations of motion we can actually use the covariant generalization of (\ref{7.27}), so that in the classical equations of motion the derivative of the metric is precisely the derivative of the metric (cf. the derivative of $y$) and not some independent variable, so that the Levi-Civita connection is given by $\Gamma^{\lambda}_{\mu\nu}=\tfrac{1}{2}\bar{g}^{\lambda\sigma} [\partial_{\mu}\bar{g}_{\nu\sigma}+\partial_{\nu}\bar{g}_{\mu\sigma}-\partial_{\sigma}\bar{g}_{\mu\nu}]$. This can be understood by noting that we can obtain the geodesic equation by varying the proper time action $\int ds [\bar{g}_{\mu\nu}(dx^{\mu}/ds)(dx^{\nu}/ds)]^{1/2}$ not with respect to $\bar{g}_{\mu\nu}$ itself but with respect to the test particle coordinate $x^{\mu}$. In this variation one precisely generates derivatives of  the metric with respect to $x^{\mu}$ and not some new independent quantity.
 
 To conclude we would like to emphasize that the treatments of the classical and quantum theories have an intrinsic difference. In the classical theory the $x=i\dot{y}$ does hold at the stationary minimum, and can even be imposed on a restricted set of variational paths in which only $y$ is varied. However,  in a path integral quantization every path must be included, including ones in which $x=i\dot{y}$  does not hold, just as we now show.
 
\section{Path integrals and the normalization of the vacuum}
\label{S8}

We are interested in the path integral approach for both the PU oscillator and the second-order plus fourth-order field theory. We treat the field theory path integral  in Sec. \ref{S12}. For the PU oscillator the Minkowski path integral  ($PI$) associated with $I_{\rm PU}$ is of the form 
\begin{align}
PI(MINK)=\int D[z]D[dz/dt]\exp\left[\frac{i}{2}\int_{-\infty}^{\infty} dt\left(\left(\frac{d^2 z}{dt^2}\right)^2-\left(\omega_1^2
+\omega_2^2\right)\left(\frac{d z}{dt}\right)^2+\omega_1^2\omega_2^2z^2\right)\right],
\label{8.1}
\end{align}
as initially integrated over paths with real $z$ and $x$. 
Since the theory is fourth order we need four pieces of information to solve the equations of motion. The pieces that are the most convenient for path integral purposes are two initial and two final conditions, hence the path integral measure is over both $z$ and $dz/dt$. However, we had noted in Sec. \ref{S5} that the PU theory is a constrained theory. Thus we must treat the $z$ and $dz/dt$ path integrations as independent.  We can do this directly as shown in the measure in (\ref{8.1}), or replace (\ref{8.1}) by 
\begin{align}
PI(MINK,z,x)=\int D[z]D[x]\exp\left[\frac{i}{2}\int_{-\infty}^{\infty} dt\left(\left(\frac{dx}{dt}\right)^2-\left(\omega_1^2
+\omega_2^2\right)x^2+\omega_1^2\omega_2^2z^2\right)\right].
\label{8.2}
\end{align}

To make the path integral converge rather than just oscillate we first use the Feynman $i\epsilon$ prescription and replace $\omega_1^2$ and $\omega_2^2$ by $\omega_1^2-i\epsilon$ and $\omega_2^2-i\epsilon$. This yields 
\begin{align}
PI(MINK,z,x)=\int D[z]D[x]\exp\left[\frac{1}{2}\int_{-\infty}^{\infty} dt\left(i\left(\frac{d x}{dt}\right)^2-i\left(\omega_1^2
+\omega_2^2\right)x^2+i\omega_1^2\omega_2^2z^2
-2\epsilon x^2+\epsilon\left(\omega_1^2+\omega_2^2\right)z^2
\right)\right],
\label{8.3}
\end{align}
as integrated over paths with real $x$ and real $z$. However, while the $x^2$ term is now damped the $z^2$ term is not. Consequently, as integrated with a real measure the path integral does not exist. Now the path integral is used to generate time-ordered Green's functions  such as $D(x)=i\langle \Omega\vert T[\phi(x)\phi(0)]\vert\Omega\rangle$ (hence the $i\epsilon$ prescription). And thus these Green's functions will not be finite, with the vacuum in which the Green's function matrix elements are evaluated thus not being normalizable. Study of the Minkowski path integral thus gives us an alternate way to determine whether or not $\langle \Omega\vert \Omega\rangle$ is finite: the path integral with a real measure either exists or does not exist.

To make (\ref{8.2}) exist we need to damp the $z^2$ term, but not modify the $x^2$ term. Thus we continue $z$ into the complex plane and replace it by $y=-iz$, while leaving $x$ real. The path integral for Minkowski time then takes the form 
\begin{align}
PI(MINK,y,x)=\int D[y]D[x]\exp\left[\frac{1}{2}\int_{-\infty}^{\infty} dt\left(i\left(\frac{d x}{dt}\right)^2-i\left(\omega_1^2
+\omega_2^2\right)x^2-i\omega_1^2\omega_2^2y^2
-2\epsilon x^2-\epsilon\left(\omega_1^2+\omega_2^2\right)y^2
\right)\right].
\label{8.4}
\end{align}
This puts us into a domain in the complex plane (known as a Stokes wedge) in which the path integral is now fully defined, and now the vacuum state is normalizable. This completely parallels the discussion of $\psi_0(z,x)$ that we gave in Secs. \ref{S5} and \ref{S7}.

\begin{figure}[htpb]
\centering
\includegraphics[scale=0.5]{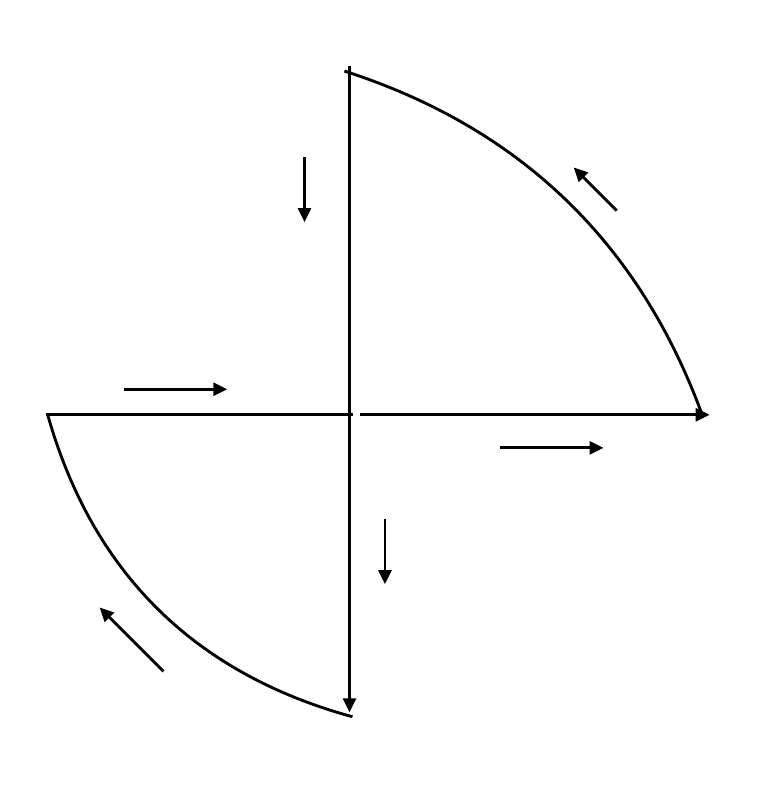}
\caption{Wick contour}
\label{wickcontour}
\end{figure}

However, our concern here could be missed in a Euclidean time path integral approach. Specifically, if we disperse in $t$ (assuming of course that we can, i.e., that the Cauchy-Riemann equations for complex $t$ are obeyed), we can, as shown in Fig. \ref{wickcontour}, write
\begin{align}
\int_{-\infty}^{\infty} +\int_{\infty}^{i\infty}+ \int_{i\infty}^{-i\infty}+ \int_{-i\infty}^{-\infty}={\rm pole~terms~plus ~cut~contributions},
\label{8.5}
\end{align}
i.e., along the real axis, then upper-half-plane quarter circle, then down the imaginary axis, and then lower-half-plane quarter circle.  Assuming no pole, cut  or circle contributions, and on setting $\tau=it$ and letting $I$ denote the action,  from (\ref{8.1}) and (\ref{8.2}) we obtain 
\begin{align}
I(MINK,z,x)&\equiv \int_{-\infty}^{\infty}idt\equiv -\int_{i\infty}^{-i\infty}idt=-\int_{-\infty}^{\infty} d\tau \equiv I(EUCL,z,x),
\nonumber\\
PI(EUCL,z,x)&=\int D[z]D[dz/d\tau]\exp\left[-\frac{1}{2}\int_{-\infty}^{\infty} d\tau\left(\left(\frac{d^2 z}{d\tau^2}\right)^2+\left(\omega_1^2
+\omega_2^2\right)\left(\frac{d z}{d\tau}\right)^2+\omega_1^2\omega_2^2z^2\right)\right]
\nonumber\\
&=\int D[z]D[x]\exp\left[-\frac{1}{2}\int_{-\infty}^{\infty} d\tau\left(\left(\frac{dx}{d\tau}\right)^2+\left(\omega_1^2
+\omega_2^2\right)x^2+\omega_1^2\omega_2^2z^2\right)\right]. 
\label{8.6}
\end{align}
Given the overall minus sign that multiplies the Euclidean action on every path, we see that with real $z$ and real $x=dz/d\tau$ the Euclidean path integral is well behaved. (The same is true of the analog relativistic second-order plus fourth-order scalar field theory path integral (see 
\cite{Hawking2002} and Sec. \ref{S12B}).) However, the Minkowski time path integral with a real measure is not. Thus we conclude that the pole and/or  cut and/or circle contributions are not only not ignorable, they generate an infinite contribution. Hence their contribution in a Wick rotation cannot be ignored and the Euclidean time path integral does not correctly describe the situation. 

In parallel, if we set $y=-iz$, then (\ref{8.6}) is replaced by
\begin{align}
I(MINK,y,x)&\equiv I(EUCL,y,x),
\nonumber\\
PI(EUCL,y,x)&=\int D[y]D[x]\exp\left[\frac{1}{2}\int_{-\infty}^{\infty} d\tau\left(-\left(\frac{dx}{d\tau}\right)^2-\left(\omega_1^2
+\omega_2^2\right)x^2+\omega_1^2\omega_2^2y^2\right)\right].
\label{8.7}
\end{align}
And since the $y$ and $x$ path integrations are independent, now it is the Euclidean time path integral that is not well defined. Thus with either real $z$ or real $y=-iz$, in neither case are the Minkowski time and Euclidean time path integrals simultaneously finite.

As we had noted above,  the quantum theory associated with (\ref{8.2}) has two distinct realizations, the Ostrogradski one and the Feynman $i\epsilon$ prescription. Now it cannot be the case that one and the same path integral has two totally different realizations, and so we need differentiate between them.  This is done by having them be defined with differing variables (and as we discuss in Sec. \ref{S12B} different $i\epsilon$ prescriptions). Thus for Ostrogradski realization $z$ is real, while for the Feynman $i\epsilon$ prescription $z$ is pure imaginary. These two options for $z$ then lead to two completely different theories. However in the end in the Feynman $i\epsilon$ prescription case we still end up with a real $y=-iz$ in (\ref{8.4}), just as needed to make the path integral exist.

As we discuss in Sec. \ref{S12}, these same results carry over directly to the field theory case, and thus we see that even if finite, a Euclidean time path integral approach is only valid if the vacuum state of the theory (as determined in a Minkowski time analysis) is normalizable.

\section{Interactions}
\label{S9}

In developing Wick's contraction theorem in quantum field theory one needs to put the time-ordered product of Heisenberg fields $\phi(x)$, viz.
\begin{align}
\tau(x_1,...,x_n)=\langle \Omega \vert T[\phi(x_1)...\phi(x_n)]\vert \Omega \rangle,
\label{9.1}
\end{align}
into a form that can be developed perturbatively. To this end one introduces a set of in-fields $\phi_{in}(x)$ that satisfy free field equations with Hamiltonian $H_{in}$. And one also introduces an evolution operator $U(t)$ that evolves with the interaction Hamiltonian $H_I(t)$ according to
\begin{align}
i\frac{\partial U(t)}{\partial t}=H_I(t)U(t).
\label{9.2}
\end{align}
With this $U(t)$ we can relate $\phi(x)$ and $\phi_{in}(x)$ according to
\begin{align}
\phi(\bar{x},t)=U^{-1}(t)\phi_{in}(\bar{x},t)U(t).
\label{9.3}
\end{align}
If one introduces $U(t,t^{\prime})=U(t)U^{-1}(t^{\prime})$, then  $U(t,t^{\prime})$ is given by 
\begin{align}
U(t,t^{\prime})=1-i\int_{t^{\prime}}^tdt_1H_I(t_1)U(t_1,t^{\prime})=T\left[\exp\left(-i\int_{t^{\prime}}^tdt_1H_I(t_1)\right)\right].
\label{9.4}
\end{align}
Using these relations we obtain (see e.g. \cite{Bjorken1965}) 
viz.
\begin{align}
\tau(x_1,...,x_n)=\langle \Omega \vert U^{-1}(t)T\left[\phi_{in}(x_1)...\phi_{in}(x_n)\exp\left(-i\int_{-t}^tdt_1H_I(t_1)\right)\right]U(-t)\vert \Omega \rangle.
\label{9.5}
\end{align}
The contributions due to the  $U(t)\vert \Omega \rangle$ and $\langle \Omega \vert U^{-1}(t)$ terms lead to
\begin{align}
\tau(x_1,...,x_n)&=\langle \Omega \vert T\left[\phi_{in}(x_1)...\phi_{in}(x_n)\exp\left(-i\int_{-t}^tdt_1H_I(t_1)\right)\right]\vert \Omega \rangle
\nonumber\\
&\times 
\langle \Omega \vert T\left[\exp\left(i\int_{-t}^tdt_1H_I(t_1)\right)\right]\vert \Omega \rangle.
\label{9.6}
\end{align}
After inverting the last term we obtain the standard form \cite{Bjorken1965} 
\begin{align}
\tau(x_1,...,x_n)&=\frac{\langle \Omega \vert T\left[\phi_{in}(x_1)...\phi_{in}(x_n)\exp\left(-i\int_{-t}^tdt_1H_I(t_1)\right)\right]\vert \Omega \rangle}
{\langle \Omega \vert T\left[\exp\left(-i\int_{-t}^tdt_1H_I(t_1)\right)\right]\vert \Omega \rangle}.
\label{9.7}
\end{align}

If one starts with (\ref{9.7}) it would appear that the normalization of the vacuum state is actually irrelevant since it would drop out of the ratio. And so it would not appear to matter if it did happen to be infinite. However, this is not the case since we could only go from (\ref{9.6}) to (\ref{9.7}) if $\langle \Omega \vert T\left[\exp\left(i\int_{-t}^tdt_1H_I(t_1)\right)\right]\vert \Omega \rangle$ is finite. And it would not be if the vacuum state is not normalizable. If we expand $\langle \Omega \vert T\left[\exp\left(i\int_{-t}^tdt_1H_I(t_1)\right)\right]\vert \Omega \rangle$ out as a power series in $H_I$ the first term is $\langle \Omega \vert  \Omega \rangle$ as calculated in a free theory. Thus, as we had noted in Sec. \ref{S3},  for finiteness we need this term to be finite and need the power series expansion in $H_I$ to be renormalizable in order for $\langle \Omega \vert T\left[\exp\left(i\int_{-t}^tdt_1H_I(t_1)\right)\right]\vert \Omega \rangle$ to be finite.  However, for a nonnormalizable vacuum the standard Wick expansion and Feynman rules are not valid. Since this concern is of relevance to radiative corrections to Einstein gravity we return to this point in Secs. \ref{S11} and \ref{S12} below.

\section{Fermions}
\label{S10}

For fermions we have to deal with anticommutators such as
\begin{align}
bb^{\dagger}+b^{\dagger}b=1.
\label{10.1}
\end{align}
Also, because of the Pauli principle we have
\begin{align}
b^2=0,\qquad b^{\dagger 2}=0. 
\label{10.2}
\end{align}
We can represent (\ref{10.1}) and (\ref{10.2}) by matrices of the form
\begin{eqnarray}
b=\begin{pmatrix}
0& 0 \\ 
1 &0
\end{pmatrix},
\qquad 
b^{\dagger}=
\begin{pmatrix}
0& 1 \\
 0 &0
 \end{pmatrix}.
\label{10.3}
\end{eqnarray}
Thus, unlike the infinite-dimensional matrix representation of the bosonic $a$ and $a^{\dagger}$ that obey $aa^{\dagger}-a^{\dagger}a=1$, the fermionic $b$ and $b^{\dagger}$ matrices are finite dimensional. Thus with a finite number of degrees of freedom, the fermion vacuum that obeys $b\vert \Omega\rangle=0$ has a finite $\langle \Omega\vert \Omega \rangle$ norm.

\section{Implications for radiative corrections in quantum Einstein gravity}
\label{S11}

As a quantum theory the standard second-order-derivative Einstein gravitational theory with its $1/k^2$ propagator is not renormalizable.  Since radiative graviton loops generate higher-derivative gravity terms, one can construct a candidate theory of quantum gravity by augmenting the Einstein Ricci scalar action with a term that is quadratic in the Ricci scalar. This gives a 
much-studied \cite{footnote6a} quantum gravity  action of the generic form
\begin{eqnarray}
I_{\rm GRAV}=\int d^4x(-g)^{1/2}\left[6M^2R^{\alpha}_{~\alpha}+(R^{\alpha}_{~\alpha})^2\right],
\label{11.1}
\end{eqnarray}
and can be considered to be an ultraviolet completion of Einstein gravity (see e.g. \cite{Donaghue2022}).
This same action also appears in Starobinsky's inflationary universe model \cite{Starobinsky1979}.

On adding on a matter source with energy-momentum tensor $T_{\mu\nu}$, variation  of this action with respect to the metric generates a gravitational equation of motion of the form
\begin{eqnarray}
-6M^2G^{\mu\nu}+V^{\mu\nu}=-\frac{1}{2}T^{\mu\nu}.
\label{11.2}
\end{eqnarray}
Here $G_{\mu\nu}$ is the Einstein tensor and $V_{\mu\nu}$ may for instance be found in \cite{Mannheim2006}, with these various terms being of the form 
\begin{eqnarray}
G^{\mu\nu}&=&R^{\mu\nu}-\frac{1}{2}g^{\mu\nu}g^{\alpha\beta}R_{\alpha\beta},
\nonumber\\
V^{\mu \nu}&=&
2g^{\mu\nu}\nabla_{\beta}\nabla^{\beta}R^{\alpha}_{~\alpha}                                             
-2\nabla^{\nu}\nabla^{\mu}R^{\alpha}_{~\alpha}                          
-2 R^{\alpha}_{~\alpha}R^{\mu\nu}                              
+\frac{1}{2}g^{\mu\nu}(R^{\alpha}_{~\alpha})^2.
\label{11.3}
\end{eqnarray}                                 
If we now linearize about  flat spacetime with background metric $\eta_{\mu\nu}$ and fluctuation  metric $g_{\mu\nu}=\eta_{\mu\nu}+h_{\mu\nu}$, to first perturbative order we obtain 
\begin{eqnarray}
\delta G_{\mu\nu}&=&\frac{1}{2}\left(\partial_{\alpha}\partial^{\alpha}h_{\mu\nu}-\partial_{\mu}\partial^{\alpha}h_{\alpha\nu}-\partial_{\nu}\partial^{\alpha}h_{\alpha\mu}+\partial_{\mu}\partial_{\nu}h\right)-\frac{1}{2}\eta_{\mu\nu}\left(\partial_{\alpha}\partial^{\alpha}h-\partial^{\alpha}\partial^{\beta}h_{\alpha\beta}\right),
\nonumber\\
\delta V_{\mu\nu}&=&[2\eta_{\mu\nu}\partial_{\alpha}\partial^{\alpha} -2\partial_{\mu}\partial_{\nu}]
[\partial_{\beta}\partial^{\beta}h-\partial_{\lambda}\partial_{\kappa}h^{\lambda\kappa}],
\label{11.4}
\end{eqnarray}                                 
where $h=\eta^{\mu\nu}h_{\mu\nu}$. On taking the trace of the linearized fluctuation equation around a flat background  we obtain 
\begin{eqnarray}
[M^2+\partial_{\beta}\partial^{\beta}]\left(\partial_{\lambda}\partial^{\lambda}h-\partial_{\kappa}\partial_{\lambda}h^{\kappa\lambda}\right)=-\frac{1}{12}\eta^{\mu\nu}\delta T_{\mu\nu}.
\label{11.5}
\end{eqnarray}
In the convenient transverse gauge where $\partial_{\mu}h^{\mu\nu}=0$, the propagator for $h$ is given by
\begin{eqnarray}
D(h,k^2)=-\frac{1}{k^2(k^2-M^2)}=\frac{1}{M^2}\left(\frac{1}{k^2}-\frac{1}{k^2-M^2}\right).
\label{11.6}
\end{eqnarray}
As we see, in this case the $1/k^2$ graviton propagator for $h$ that would be associated with the Einstein tensor $\delta G_{\mu\nu}$ alone  is replaced by a $D(h,k^2)=[1/k^2-1/(k^2-M^2)]/M^2$ propagator. And now the leading behavior at large momenta is $-1/k^4$. In consequence, the theory is  thought to be renormalizable \cite{Stelle1977,Stelle1978}. But since  $\langle \Omega\vert\Omega \rangle$ is not finite the proof of renormalizability has a flaw in it. Fortunately, the flaw is not fatal, and we rectify it below.

We recognize $D(h,k^2)$ as being of the same form as the second-order plus fourth-order scalar field theory propagator that was given in (\ref{4.2}), with $\phi$ being replaced by $h$ and with $M_1^2=M^2$, $M_2^2=0$. We can thus give $h$ an equivalent effective action of the form
\begin{eqnarray}
I_h&=&\frac{1}{2}\int d^4x\bigg{[}\partial_{\mu}\partial_{\nu}h\partial^{\mu}
\partial^{\nu}h-M^2\partial_{\mu}h\partial^{\mu}h\bigg{]}.
\label{11.7}
\end{eqnarray}
The action given in (\ref{11.7})  thus shares the same vacuum state normalization and negative norm challenges as the scalar field action given in (\ref{4.1}).

Thus if, as is conventional, we take $h$ to be Hermitian we would immediately encounter the negative-norm problem associated with the relative minus sign in (\ref{11.6}). However, since $M^2$ is Planck scale in magnitude, this difficulty can be postponed until observations can reach that energy scale. However, the lack of normalizabilty of the vacuum state has consequences at all energies and cannot be postponed at all. Specifically, with $\langle \Omega\vert\Omega \rangle$ being infinite we cannot even identify the propagator as $i\langle \Omega\vert T[h(x)h(0)]\vert \Omega \rangle$ since in analog to (\ref{1.8}) it will obey
\begin{align}
&(\partial_t^2-\bar{\nabla}^2)(\partial_t^2-\bar{\nabla}^2+M^2)D(h,x)=-\langle \Omega\vert\Omega \rangle\delta^4(x).
\label{11.8}
\end{align}
Consequently, we cannot make the standard Wick contraction expansion. And thus both the Feynman rules that are used presupposing that $\langle \Omega\vert\Omega \rangle$ is finite, and the renormalizability that is thought to then follow from them are therefore not valid. Additionally, with $\langle \Omega\vert\Omega \rangle$ being infinite, we cannot treat the Einstein theory with its $1/k^2$ propagator as an effective field theory that holds for momenta that obey $k^2 \ll M^2$.

However,  as noted above, we can resolve all of these concerns by dropping the requirement that $h$ be Hermitian, and set it equal to $i\bar{h}$. Then, with the theory being recognized as a $PT$ theory, vacuum state normalization and negative-norm problems are resolved and the theory is consistent. Moreover,  the propagator is given by $-i\langle \Omega^{L}\vert T[\bar{h}(x)\bar{h}(0)]\vert \Omega^R \rangle$ (corresponding to $+i\langle \Omega^{L}\vert T[h(x)h(0)]\vert \Omega^R \rangle$). And with the propagator still being given by (\ref{11.6}) as it satisfies $(\partial_t^2-\bar{\nabla}^2+M^2)(\partial_t^2-\bar{\nabla}^2)[-i\langle \Omega^{L}\vert T[\bar{h}(x)\bar{h}(0)]\vert \Omega^R \rangle]=-\delta^4(x)$, all the steps needed to prove renormalizability are now valid. At this point the only concern is that even though the $M^2$ field  now has positive norm, it still remains in the spectrum and would eventually have to be observed. However, with this massive field and with the continuation into the complex plane we can now recognize the second-order-derivative plus fourth-order-derivative gravitational theory as a bona fide ultraviolet completion of Einstein gravity.

As can be seen from (\ref{11.7}), the only reason that there is an $M^2$ term at all is because we are considering an action that has both second-order and fourth-order terms. With a pure fourth-order theory there would be no dimensionful parameter in the action and the theory would be scale invariant. If like the gauge theories of $SU(3)\times SU(2)\times U(1)$ this scale symmetry is also local, we would be led to conformal gravity, a metric theory of gravity in which the action is left invariant under local changes of the metric of the form $g_{\mu\nu}(x)\rightarrow e^{2\alpha(x)}g_{\mu\nu}(x)$, where $\alpha(x)$ is a local function of the coordinates. The conformal gravity theory has been advocated and explored in \cite {Mannheim2006,Mannheim2017} and references therein. And in  \cite{hooft2015local} 't Hooft has also argued that there should be an underlying local conformal symmetry in nature. 

In the conformal gravity theory an action that is to be a polynomial function of the metric has the unique form 
\begin{eqnarray}
I_{\rm W}=-\alpha_g\int d^4x\, (-g)^{1/2}C_{\lambda\mu\nu\kappa}
C^{\lambda\mu\nu\kappa}
\equiv -2\alpha_g\int d^4x\, (-g)^{1/2}\left[R_{\mu\kappa}R^{\mu\kappa}-\frac{1}{3} (R^{\alpha}_{~\alpha})^2\right],
\label{11.9}
\end{eqnarray}
where $\alpha_g$ is a dimensionless  gravitational coupling constant, and $C_{\lambda\mu\nu\kappa}$ is the conformal Weyl tensor. The perturbative propagator has a $-1/k^4$ behavior at all $k^2$, and with its large $k^2$ behavior the theory is renormalizable \cite{Fradkin1985}. With a $-1/k^4$  propagator it would initially appear that there would be two massless particles at $k^2=0$. However, we cannot use the partial fraction decomposition given in (\ref{11.6}) as a guide since its $1/M^2$ prefactor is singular in the $M^2\rightarrow 0$ limit. Because of this singular behavior the $M^2=0$ Hamiltonian becomes of nondiagonalizable Jordan-block form and only has one massless eigenstate, with the other would-be massless eigenstate becoming nonstationary \cite{Bender2008b}. (This lack of diagonalizability can  also be seen from (\ref{7.19}), since the diagonalizing $e^{-Q/2}$ operator becomes singular in the $\omega_1	=\omega_2$ limit.) As is typical of Jordan-block Hamiltonians, with the theory being Jordan block there are now zero norm states. For constructing viable Hilbert spaces they are just as acceptable as positive norm ones.

To understand why we have lost an eigenstate let us consider two positive frequency mode solutions to
\begin{eqnarray}
[M^2+\partial_{\beta}\partial^{\beta}]\partial_{\lambda}\partial^{\lambda}h=0,
\label{11.10}
\end{eqnarray}
viz. modes with time dependence $e^{-i\omega_1t}$ and $e^{-i\omega_2 t}$, where $\omega_1=|\bar{k}|$, $\omega_2=+(\bar{k}^2+M^2)^{1/2}$. These modes are solutions to  the $M^2\neq 0$ wave equation  associated with (\ref{11.5}) (as are the two negative frequency solutions with time dependence $e^{i\omega_1t}$ and $e^{i\omega_2 t}$.) If we now let $M^2$ go to zero the two positive frequency mode solutions  become equal and we only have one positive frequency mode solution. However, we cannot lose any solutions to the differential equation given in (\ref{11.10}) by setting $M^2=0$ in it since the wave equation is still fourth order. To find the other solution in the limit we consider
\begin{eqnarray} 
\lim_{M^2\rightarrow 0}\frac{(e^{-i\omega_1t}-e^{-i\omega_2t})}{M^2}=\frac{ite^{-i\omega_1t}}{2|\bar{k}|}.
\label{11.11}
\end{eqnarray}
As we see, this second solution is not stationary and is thus not an eigenstate of the Hamiltonian.

Since we have lost a massless eigenstate  the propagator should be constructed not as the $M^2\rightarrow 0$ limit of (\ref{11.6}) but as the manifestly ghost-free limit
\begin{eqnarray}
-\frac{1}{(k^2+i\epsilon)^2}=-\lim_{M^2\rightarrow 0} \frac{d}{dM^2}
\left(\frac{1}{k^2-M^2+i\epsilon}\right),
\label{11.12}
\end{eqnarray}
a limit that shows that there is only one $k^2=0$ pole not two. With the Hamiltonian not being diagonalizable, it could not be Hermitian. It does however have a $PT$ symmetry, with is ground state being normalizable. Conformal gravity is thus a fully consistent theory of quantum gravity, one which despite its fourth-order character only possesses one massless particle, not two  \cite{footnote4}.

To understand why both second-order plus fourth-order theories and pure fourth-order theories must be ghost free consider the Dirac action for a fermion coupled to a background geometry of the form
\begin{eqnarray}
I_{\rm D}=\int d^4x(-g)^{1/2}\left[i\bar{\psi}\gamma^{c}V^{\mu}_c(x)(\partial_{\mu}+\Gamma_{\mu}(x))\psi -M\bar{\psi}(x)\psi(x)\right].
\label{11.13}
\end{eqnarray}
Here the $V^{\mu}_a(x)$ are vierbeins, $\Gamma_{\mu}(x)=-(1/8)[\gamma_a,\gamma_b](V^b_{\nu}(x)\partial_{\mu}V^{a\nu}(x)+V^b_{\lambda}(x)\Gamma^{\lambda}_{\nu\mu}(x)V^{a\nu}(x))$ is the spin connection, and $\Gamma^{\lambda}_{\nu\mu}(x)$ is the geometric Levi-Civita connection.  We introduce the path integral $\int D[\psi]D[\bar{\psi}]\exp(iI_{\rm D})=\exp(iI_{\rm EFF})$. With $I_{\rm D}$ being linear in both $\psi$ and $\bar{\psi}$ the fermion path integration can be performed analytically, to thereby yield an effective action $I_{\rm EFF}$ whose leading term is of the form \cite{tHooft2010a}, \cite{Mannheim2017}
\begin{eqnarray}
I_{\rm EFF}&=&\int d^4x(-g)^{1/2}C\bigg{[}\frac{1}{20}\left[R_{\mu\nu}R^{\mu\nu}-\frac{1}{3}(R^{\alpha}_{~\alpha})^2\right]
-M^4+\frac{1}{6}M^2R^{\alpha}_{~\alpha}\bigg{]},
\label{11.14}
\end{eqnarray}
where $C$ is a log divergent constant. We recognize $I_{\rm EFF}$ as containing none other than second-order and fourth-order gravitational terms. Now the $I_D$ action is a completely standard action of a fermion coupled to a background gravitational field, and as such it is not only ghost free, it would remain so if the fermion is given some internal quantum numbers and coupled to some gauge fields. Since the fermion path integral is equivalent to a one loop Feynman diagram and since one cannot change the signature of a Hilbert space in perturbation theory, higher-derivative gravity must be ghost free too. And if it were not, then the standard $SU(3)\times SU(2)\times U(1)$ model would not remain unitary when coupled to gravity. 

Thus despite the negative-norm issue, higher-derivative gravity cannot contain any negative-norm states. And for the standard model not to be destabilized when coupled to gravity, the subsequent $g_{\mu\nu}$ path integration must be conducted with a measure $D[g_{\mu\nu}]$ that is continued into the complex plane, just as found with $PT$ theory.
Since, as noted in Sec. \ref{S8}, this continuation is required by the $i\epsilon$ prescription, we now comment on this prescription in  more detail.

\section{$PT$ symmetry and the $i\epsilon$ prescription}
\label{S12}

\subsection{Propagators}
\label{S12A}

The $i\epsilon$ prescription is central to both propagators and path integrals, and discussion of its role provides a quick explanation of our results.  For the second-order plus fourth-order theory discussed in Sec. \ref{S4} the  identification of the $D(k)$ propagator in (\ref{4.2}) as
\begin{align}
&D(k)=-\frac{1}{(k^2-M_1^2)(k^2-M_2^2)}
=- \frac{1}{(M_1^2-M_2^2)}\left[\frac{1}{(k^2-M_1^2)}-\frac{1}{(k^2-M_2^2)}\right]
\label{12.1}
\end{align}
is only formal since $D(k)$ is singular. To give it a meaning we need to define it via a contour integral and specify the appropriate complex $k_0$ plane contour. If, as is conventional,  we take all of the operators in $H=\int d^3x T_{00}$  to be Hermitian, where as given in (\ref{4.3})
\begin{align}
T_{00}&=\tfrac{1}{2}\pi_{00}^2+\pi_{0}\dot{\phi}+\tfrac{1}{2}(M_1^2+M_2^2)\dot{\phi}^2-\tfrac{1}{2}M_1^2M_2^2\phi^2
-\tfrac{1}{2}\pi_{ij}\pi^{ij}+\tfrac{1}{2}(M_1^2+M_2^2)\phi_{,i}\phi^{,i},
\label{12.2}
\end{align}
we immediately find that because of the $-(1/2)M_1^2M_2^2\phi^2$ term the Hamiltonian $H$ is unbounded from below, the Ostrogradski instability that is characteristic of higher-derivative theories. Now the standard Feynman contour $k^2+i\epsilon$ prescription with
\begin{align}
&D(k)=-\frac{1}{(k^2-M_1^2+i\epsilon)(k^2-M_2^2+i\epsilon)}
=- \frac{1}{(M_1^2-M_2^2)}\left[\frac{1}{(k^2-M_1^2+i\epsilon)}-\frac{1}{(k^2-M_2^2+i\epsilon)}\right]
\label{12.3}
\end{align}
is chosen so that positive energy states propagate forward in time (viz.  $\omega_1(\bar{k})=+(\bar{k}^2+M_1^2)^{1/2}$, $\omega_2(\bar{k})=+(\bar{k}^2+M_2^2)^{1/2}$ located below the real $k_0$ axis), while negative energy states propagate backwards ($-\omega_1(\bar{k})$, $-\omega_2(\bar{k})$ above the real $k_0$ axis); with this corresponding to an energy spectrum that is bounded from below while leading to negative residues that appear to correspond to states of negative norm. In contrast, having an energy spectrum that is unbounded from below  would instead entail that  negative energies in one of sectors ($\omega_2(\bar{k})$ say) are propagating forward in time (-$\omega_2(\bar{k})$ below the real $k_0$ axis), and positive energies are propagating backward in time ($\omega_2(\bar{k})$ above the real $k_0$ axis), and would require using an unconventional $i\epsilon$ prescription \cite{Bender2008b} of the form
\begin{align}
&D(k)=-\frac{1}{(k^2-M_1^2+i\epsilon)(k^2-M_2^2-i\epsilon)}
=- \frac{1}{(M_1^2-M_2^2)}\left[\frac{1}{(k^2-M_1^2+i\epsilon)}-\frac{1}{(k^2-M_2^2-i\epsilon)}\right].
\label{12.4}
\end{align}
While this possibility is unacceptable physically, it does have the feature that because of the way the singularities are traversed all pole residues are positive. Thus the two options are: bounded energies and negative residues, or unbounded energies and positive residues. These two realizations are inequivalent and correspond to different Feynman contours and different Hilbert spaces. As discussed in Sec. \ref{S5} and in \cite{Bender2008b} the first option corresponds to working in a Hilbert space in which $a_2$ effects $a_2\vert \Omega\rangle=0$,  while the latter corresponds to working in a Hilbert space in which $a^{\dagger}_2\vert \Omega\rangle=0$. Thus in no Hilbert space do we have both negative energies and negative residues. Thus if we work in the Hilbert space in which $a_2\vert \Omega \rangle=0$ (which we do in this paper) and use the standard Feynman $i\epsilon$ prescription given in (\ref{12.3}),  our results will not be affected by the Ostrogradski instability at all. Nor will it affect the continuation into the complex plane that resolves the negative-norm and vacuum-normalization issues.

\subsection{Path integrals}
\label{S12B}

In analog to the quantum-mechanical (\ref{8.2}) and the field theory study given in \cite{Hawking2002} the Minkowski path integral associated with the field theory action given in (\ref{4.1}) is of the form
\begin{align}
PI(MINK)=\int D[\phi]D[\sigma_{\mu}]\exp\left[\frac{i}{2}\int_{-\infty}^{\infty} d^4x\left[\partial_{\nu}\sigma_{\mu}\partial^{\nu}\sigma^{\mu}-\left(M_1^2+M_2^2\right)\sigma_{\mu}\sigma^{\mu}+M_1^2M_2^2\phi^2\right]\right],
\label{12.5}
\end{align}
where $\sigma_{\mu}=\partial_{\mu}\phi$. 
In order to damp out oscillations we choose the Feynman $i\epsilon$ prescription in which we replace $M_1^2$ and $M_2^2$ by $M_1^2-i\epsilon$, $M_2^2-i\epsilon$. For the path integral this yields
\begin{align}
PI(MINK)&=\int D[\phi]D[\sigma_{\mu}]\exp\bigg{[}\frac{1}{2}\int_{-\infty}^{\infty} d^4x\big{[}i\partial_{\nu}\sigma_{\mu}\partial^{\nu}\sigma^{\mu}-i\left(M_1^2+M_2^2\right)\sigma_{\mu}\sigma^{\mu}+iM_1^2M_2^2\phi^2
\nonumber\\
&-2\epsilon\sigma_{\mu}\sigma^{\mu}+(M_1^2+M_2^2)\epsilon\phi^2\big{]}\bigg{]}.
\label{12.6}
\end{align}
With $\phi$ and $\sigma_{\mu}$ being taken to be real and with $\sigma_{\mu}\sigma^{\mu}$ being taken to be timelike on every path, the $\sigma_{\mu}$ path integration is damped but  the $\phi$ path integration is not. 

For the unconventional $i\epsilon$ prescription in which we replace $M_1^2$ and $M_2^2$ by $M_1^2-i\epsilon$, $M_2^2+i\epsilon$ the path integral takes the form
\begin{align}
PI(MINK)&=\int D[\phi]D[\sigma_{\mu}]\exp\bigg{[}\frac{1}{2}\int_{-\infty}^{\infty} d^4x\big{[}i\partial_{\nu}\sigma_{\mu}\partial^{\nu}\sigma^{\mu}-i\left(M_1^2+M_2^2\right)\sigma_{\mu}\sigma^{\mu}+iM_1^2M_2^2\phi^2-(M_1^2-M_2^2)\epsilon\phi^2\big{]}\bigg{]},
\label{12.7}
\end{align}
and has no damping on the $\sigma_{\mu}$ path integration at all. The unconventional $i\epsilon$ prescription for the Feynman contour that leads to an unbounded from below energy spectrum thus cannot be associated with a well-defined path integral, and we cannot consider it further.

Thus the only Feynman $i\epsilon$ prescription that can be relevant is the standard one with $M_1^2-i\epsilon$, $M_2^2-i\epsilon$. However even with this choice the $\phi$ path integration is not damped if $\phi$ is real. It becomes damped if we do not require $\phi$ to be real, but instead take it to be pure imaginary (though $({\rm Im}[\phi])^2> ({\rm Re}[\phi])^2$ would suffice). With $\bar{\phi}=-i\phi$ we replace (\ref{12.6}) by
\begin{align}
PI(MINK)&=\int D[\bar{\phi}]D[\sigma_{\mu}]\exp\bigg{[}\frac{1}{2}\int_{-\infty}^{\infty} d^4x\big{[}i\partial_{\nu}\sigma_{\mu}\partial^{\nu}\sigma^{\mu}-i\left(M_1^2+M_2^2\right)\sigma_{\mu}\sigma^{\mu}-iM_1^2M_2^2\bar{\phi}^2
\nonumber\\
&-2\epsilon\sigma_{\mu}\sigma^{\mu}-(M_1^2+M_2^2)\epsilon\bar{\phi}^2\big{]}\bigg{]}.
\label{12.8}
\end{align}
With $\bar{\phi}$ and $\sigma_{\mu}$ being taken to be real and with $\sigma_{\mu}\sigma^{\mu}$ being taken to be timelike on every path, the path integral is now well defined and the theory is consistent. (In a study of quantum gravity 't Hooft 
\cite{tHooft2011} has also suggested that the path integral measure be continued into the complex domain.)

In classical mechanics Poisson bracket relations are unaffected by symplectic transformations, while in quantum mechanics commutation relations are unaffected by similarity transformations. With path integrals being based on classical fields, the utility of making a symplectic transformation on the classical fields is that it can take a badly-behaved path integral with a real measure  into a complex domain for the measure known as a Stokes wedge in which the path integral then is well behaved. Analogously, as described in Sec. \ref{S7}, a similarity transformation can take us into a complex domain in which nonnormalizable quantum wave functions  become normalizable. In both the classical and quantum cases we thus seek appropriate complex domains that have good boundary behavior.

While one is always free to make symplectic transformations or similarity transformations, for standard positive Dirac norm Hermitian theories with energies  that are bounded from below, these transformations contain no new information. For such Hermitian theories the quantum fields are Hermitian, the classical fields that appear in the path integral are their real eigenvalues, the path integral with a real measure and Feynman $i\epsilon$ prescription exists, the Dirac norm of the vacuum is finite, and the path integral can be associated with matrix elements of the form $i\langle \Omega \vert T[\phi(x)\phi(0)]\vert \Omega\rangle$.

However, it can happen that the path integral with real measure does not exist, but that it does exist for some appropriately chosen complex domain for the measure. In that case the Dirac norm of the vacuum is not finite. Also the Hamiltonian will not be Hermitian, but if all the poles in the propagator are real,  the Hamiltonian will instead be $PT$ symmetric. In that case the propagator is given by $i\langle \Omega^L \vert T[(i\bar{\phi}(x))(i\bar{\phi}(0))]\vert \Omega^R\rangle
=-i\langle \Omega^L \vert T[\bar{\phi}(x)\bar{\phi}(0)]\vert \Omega^R\rangle$, where $\phi$ is transformed into $-i\phi=\bar{\phi}$. And the vacuum norm is given by $\langle \Omega^L \vert \Omega^R\rangle$, a norm that is finite, positive definite and time independent. Thus when we find for the second-order-derivative  plus fourth-order-derivative quantum field theory that propagator pole residues are negative or that the path integral with real measure does not exist, it does not mean that the theory does not exist, but that it has to be formulated in a Hilbert space with an inner product other than the standard Dirac one. 

Since for the second-order-derivative plus fourth-order-derivative theory  the vacuum Dirac norm is not finite,  the standard derivation of the Feynman rules from the Wick contraction procedure is invalid, as is then the renormalizabilty that would follow from these Feynman rules. Interestingly, since the pole structure of the propagator is not affected by a similarity transformation, even after the transformation we can still use the propagator given in (\ref{12.3}) with its $-1/k^4$ short-distance behavior. The only thing that changes is the quantum field matrix element that we identify it with. This leads to the same Feynman rules, only now validly derived from a Wick contraction procedure in which everything is finite, to thus provide for an a posteriori derivation  of renormalizability, while deriving it now in a Hilbert space that possesses no states with negative norm. The renormalizability and unitarity of the second-order plus fourth-order  field theory is thus established,  just as needed for a consistent theory of quantum gravity. 

\subsection{Complex correspondence principle}
\label{S12C}

It is of interest to discuss our findings from the perspective of the correspondence principle. While the correspondence principle is not in and of itself a law of nature, it has proven to be very useful in providing a bridge between classical physics and quantum physics, though it is not a complete guide as to what quantum theories might be permissible. Since central to the use of the correspondence principle is that its starting point is based on a classical physics with real quantities, and since $PT$ theory involves a continuation into the complex plane, we need to reconcile the two approaches,  a concern that has for instance been raised in \cite{Woodard2015}, and especially so since in experiments one only measures real quantities.

As introduced into physics there are two versions of the correspondence principle. The first one is due to Bohr and was developed in the old quantum theory era, with the second form emerging as part of the development of quantum mechanics in 1925. The Bohr approach was to show that in the limit of large quantum numbers solutions to the quantum Bohr atom approached the solutions to the corresponding classical system. The quantum-mechanical formulation  was to replace classical Poisson brackets by quantum commutators (viz. canonical quantization). Central to this latter approach is  that the eigenfunctions of the quantum Hamiltonian be complete  and normalizable on the real axis, that the quantum Hamiltonian be self-adjoint when acting on these eigenfunctions,  and that the quantum Hamiltonian be Hermitian so that its energy eigenvalues be real. To connect with classical physics  the quantum position and momentum operators that appear in the quantum Hamiltonian have to be self-adjoint when acting on the eigenfunctions of the quantum Hamiltonian, so that their position and momentum eigenvalues would indeed be real, with wave functions being given by $\psi(x)=\langle x\vert \psi\rangle$ with real $x$ (viz. the probability of finding an eigenstate of the position operator in an eigenstate of the Hamiltonian). Quantum operators that satisfy all of these criteria are observables.

With the advent of quantum field theory the bridge between classical physics and quantum physics had to be modified. While one can still determine the form of quantum commutators by a canonical quantization of Poisson brackets, the classical limit is obtained not by looking at the $\hbar\rightarrow 0$ limit, but by taking matrix elements of the quantum fields in states with an indefinite number of field quanta (a large number limit, just like Bohr's formulation of the correspondence principle). If the quantum theory is renormalizable the output matrix elements will obey the same equations as the classical field equations, albeit with renormalized, and thus $\hbar$-dependent  masses and coupling constants. Thus if for instance we were to define  the classical electrodynamics limit of quantum electrodynamics as the $\hbar \rightarrow 0$ limit we would have to use the bare charge, which we could not do since it is infinite. Using the renormalized charge would give a finite classical electrodynamics but not an $\hbar$-independent one. The classical limit is a c-number limit, not an $\hbar=0$ limit.

In regard first to the Bohr large quantum number approach,  it has actually been shown that $PT$ theory is compatible with it, with the old quantum theory correspondence principle being extended to the complex domain \cite{Bender2010}. Moreover, not only are complex plane quantum-mechanical probabilities found to approach complex plane classical-mechanical probabilities in the large quantum number limit, they do so with more precision than in a typical real variable case.

A complex extension of the quantum-mechanical correspondence principle also exists, and may be found in the second-order-derivative plus fourth-order-derivative field theory case. Specifically, first with real variables  the nonrelativistic PU Hamiltonian $H_{\rm PU}$ given in (\ref{5.2}) was constructed by obtaining a closed real variable Poisson bracket algebra for the classical Hamiltonian and the operators in it \cite{Mannheim2000,Mannheim2005}, and then replacing it by a closed quantum commutation algebra  involving the quantum Hamiltonian  and the quantum commutators given in (\ref{5.3}). The relativistic field theory generalization presented in Sec. \ref{S4} was then given in \cite{Bender2008b}. Thus at this point we are explicitly following the correspondence principle in its standard real variable formulation. To then ascertain whether the quantum operators can be considered to be observables we need to determine whether  are they are self-adjoint, and if so whether then their eigenvalues are real.

This procedure can actually be carried out as is in the real variable Ostrogradski realization of the second-order-derivative plus fourth-order-derivative theory. As noted in \cite{Woodard2015} and in Sec. \ref{S5} above, in this case the wave functions of the quantum $H_{\rm PU}$ Hamiltonian are convergent so that $H_{PU}$ is self-adjoint (and by the general analysis given above so is $H_S$), and all of the $H_{\rm PU}$ and $H_S$ eigenvalues are real. Thus both $H_{\rm PU}$ and $H_S$  are observables, and the correspondence principle in its real variable formulation is obeyed. However, while real, in both cases the energy spectrum is unacceptably unbounded from below. Now if at this point we were to  stop and simply abandon the second-order-derivative plus fourth-order-derivative approach to quantum gravity because of its Ostrogradski instability, we would still have a problem. Specifically, we would still have to deal with the problem identified in (\ref{11.14}), namely that a fermion path integration in the presence of a  gravitational field generates higher-derivative terms anyway. Thus we cannot ignore them even if we wanted to. Thus if we stay with real variables  we would have to conclude that the standard $SU(3)\times SU(2)\times U(1)$ model of particle physics  is rendered unstable when coupled to gravity.  Moreover, even if we use the standard Feynman $i\epsilon$ prescription realization of the second-order-derivative plus fourth-order-derivative theory but keep everything real, the quantum Hamiltonian would still not be self-adjoint, the Dirac norm of the vacuum would be infinite, there would be states of negative Dirac norm in the Hilbert space, and the perturbative Wick contraction procedure would lead to Feynman rules with infinite Green's functions. So again we would not be able to make sense of (\ref{11.14}), and the standard $SU(3)\times SU(2)\times U(1)$ model of particle physics would lose unitarity when coupled to gravity.

So no matter what we do we will have to deal with higher-derivative gravity even if only because of an $SU(3)\times SU(2)\times U(1)$  fermion path integration. Moreover, this concern will continue to exist even if we base quantum gravity on string theory or loop quantum gravity. To actually deal with these higher-derivative terms  we need to extend the correspondence principle into the complex domain. As noted above in Sec. \ref{S7A}, in classical mechanics Poisson bracket relations are unaffected by symplectic transformations, while in quantum mechanics commutation relations are unaffected by similarity transformations. Moreover, this remains true even if the transformations are complex (in (\ref{7.1}) we can set $S({\rm PU})=\exp(i\alpha p_z z)$, $S(S)=\exp(i\alpha \int d^3x\pi_0(\bar{x},t)\phi(\bar{x},t))$, with complex $\alpha$). Thus if we rotate through the same angle in both the classical symplectic transformation and the quantum similarity transformation, we will obtain a complex correspondence principle at every angle in the complex plane, with the rotated Poisson bracket algebra quantizing into the rotated commutator algebra at every angle \cite{mannheim2018antilinearity}. However, as we had noted in Sec. \ref{S7A}, in making the rotation we cross into a Stokes wedge in which the operators now are self-adjoint. However, as can be seen in (\ref{7.3}) and (\ref{7.15}), the $\bar{H}_{\rm PU}$ and  $\bar{H}_S$ Hamiltonians are not Hermitian. Nonetheless, they are $PT$ symmetric and their eigenvalues are real. The rotated Hamiltonians are thus quantum observables. Moreover, for each one of them the energy spectrum is bounded from below. We are thus in the Feynman $i\epsilon$ prescription only now with a $PT$-theory inner product that is finite, positive and time independent. Thus we extend the correspondence principle into the complex domain, and in so doing obtain a fully consistent, unitary and renormalizable quantum gravity theory.

However, despite the complex rotation, as noted in Sec. \ref{S7E},  the associated classical limit does not involve complex numbers.  For gravity we note \cite{Mannheim2017a} that if we replace the covariant classical $g_{\mu\nu}$ by $\bar{g}_{\mu\nu}=ig_{\mu\nu}$ and accordingly replace the contravariant classical $g^{\mu\nu}$ by $\bar{g}^{\mu\nu}=-ig^{\mu\nu}$ the Levi-Civita connection will remain unchanged, and thus there will be no change in geodesics. Moreover,  after transforming to the then Hermitian  quantum $\bar{g}_{\mu\nu}$ its classical eigenvalue $\bar{g}_{\mu\nu}$ will  be real, with the line element being $ds^2=\bar{g}_{\mu\nu}dx^{\mu}dx^{\nu}$, viz. the line element whose variation gives the $\bar{g}_{\mu\nu}$-based geodesic. From this line element we can derive real proper times and proper distances.  In addition, as also noted in Sec. \ref{S7E}, we can define the coupling of gravity to matter ab initio with $\bar{g}_{\mu\nu}$. Thus the output classical gravity is just as real as it needs to be. Because of  the similarity transformation into the complex plane, from the action given in (\ref{11.1}) we obtain a consistent set of Feynman rules and a bonafide ultraviolet completion of Einstein gravity. And now at energy scales way below  $M^2$ we can treat Einstein gravity as an effective theory. However, this effective theory would still have both dark matter and dark energy problems that need to be addressed. These two problems could also be resolved by bypassing second-order-derivative gravity altogether and utilizing a pure fourth-order-derivative conformal gravity theory instead \cite{footnote4}.

As well as the correspondence principle there is another bridge between classical physics and quantum physics, namely path integral quantization. In this approach quantum field theory matrix elements are constructed by a path integration over all classical paths between fixed endpoints. In this approach the relevant action is the classical one. While the  path integral ordinarily exists as is with real fields if the action is based on a second-order-derivative theory, intuition developed from study of  second-order-derivative actions does not carry over to second-order-derivative plus fourth-order-derivative actions. Specifically, if we start with the action given in (\ref{4.1}) , viz. $I_S=\frac{1}{2}\int d^4x [\partial_{\mu}\partial_{\nu}\phi\partial^{\mu}
\partial^{\nu}\phi-(M_1^2+M_2^2)\partial_{\mu}\phi\partial^{\mu}\phi
+M_1^2M_2^2\phi^2]$, unless we know that it describes a constrained system we would not know that we have to treat $\sigma_{\mu}=\partial_{\mu}\phi$ as being completely independent of $\phi$. That $I_S$ would have to describe a constrained system is because in a theory with fourth-order derivatives one needs four pieces of information not two in order to integrate equations of motion that involve second-order and fourth-order derivatives of $\phi$. Moreover, for determining the classical equations from the classical action  using a functional variation one needs to hold both $\phi$ and $\sigma_{\mu}$ fixed at both end points in order to have a well-defined variational procedure. Thus for path integrals the functional integration is done over independent $\phi$ and $\sigma_{\mu}$, just as shown in (\ref{12.6}). However, even with the Feynman $i\epsilon$ prescription (\ref{12.6}) is not damped. To damp it we have to continue $\phi$ into the complex plane. However, we would lose the damping if we were at the same time to continue $\sigma_{\mu}=\partial_{\mu}\phi$ into the complex plane as well. It is only because $\phi$ and $\sigma_{\mu}$ are independent that we are able to find a domain for the path integral measure for which the path integral actually is finite. Thus even though we may have started with (\ref{12.6}), since it is not finite it is not actually of relevance to physics. Only (\ref{12.8}) is of relevance, regardless of how we may or may not have found it, and since (\ref{12.6}) is not finite it is without content, with it being only $\bar{g}_{\mu\nu}$ that is observable and not $g_{\mu\nu}$. Thus before we can attribute any physical significance to a path integral we need to first find some domain for the measure for which the path integral actually exists. Once we have found such a domain we are then assured that the path integral will generate matrix elements of time-ordered products of fields as evaluated in an appropriate quantum Hilbert space with an inner product that is finite, positive and time independent. To actually identify the relevant Hilbert space we use the techniques of $PT$ theory.

In the path integral bridge between classical and quantum physics we need the variables on the classical side to be the eigenvalues of self-adjoint operators on the quantum side. To show that this is case  it is convenient  to formulate  the action not as function of $\bar{x}$ and $t$ but as a function of $\bar{k}$ and $t$. From the $\bar{k}$-dependent form for the Hamiltonian given in (\ref{6.5}) and though use of the Legendre transform we rewrite the scalar field action given in (\ref{4.1}) in the $\bar{k}$-dependent form 
\begin{align}
I_S&=\int dt \int d^3k\bigg{[}p_x(\bar{k},t)\dot{x}(\bar{k},t)+p_z(\bar{k},t)\dot{z}(\bar{k},t)-\frac{p_x^2(\bar{k},t)}{2}-p_z(\bar{k},t)x(\bar{k},t)
\nonumber\\
&-\frac{1}{2}\left[\omega_1^2(\bar{k})+\omega_2^2(\bar{k}) \right]x^2(\bar{k},t)+\frac{1}{2}\omega_1^2(\bar{k})\omega_2^2(\bar{k})z^2(\bar{k},t)\bigg{]}.
\label{12.9}
\end{align}
With $\dot{z}(\bar{k},t)=x(\bar{k},t)$, $p_x(\bar{k},t)=\dot{x}(\bar{k},t)$, for the action and path integral  we obtain
\begin{align}
I_S&=\int dt \int d^3k\bigg{[}\frac{\dot{x}^2(\bar{k},t)}{2}-\frac{1}{2}\left[\omega_1^2(\bar{k})+\omega_2^2(\bar{k}) \right]x^2(\bar{k},t)+\frac{1}{2}\omega_1^2(\bar{k})\omega_2^2(\bar{k})z^2(\bar{k},t)\bigg{]},
\label{12.10}
\end{align}
\begin{align}
PI(MINK)=\int D[z]D[x]\exp\bigg{[}i\int dt d^3k\left[\frac{\dot{x}^2(\bar{k},t)}{2}-\frac{1}{2}\left[\omega_1^2(\bar{k})+\omega_2^2(\bar{k}) \right]x^2(\bar{k},t)+\frac{1}{2}\omega_1^2(\bar{k})\omega_2^2(\bar{k})z^2(\bar{k},t)\right]\bigg{]}.
\label{12.11}
\end{align}
To dampen the path integral we first replace $\omega_1^2(\bar{k})$ and  $\omega_2^2(\bar{k})$ by $\omega_1^2(\bar{k})-i\epsilon$ and  $\omega_2^2(\bar{k})-i\epsilon$ to obtain
\begin{align}
PI(MINK)&=\int D[z]D[x]\exp\bigg{[}\int dt d^3k\bigg{[}\frac{i\dot{x}^2(\bar{k},t)}{2}-\frac{i}{2}\left[\omega_1^2(\bar{k})+\omega_2^2(\bar{k}) \right]x^2(\bar{k},t)+\frac{i}{2}\omega_1^2(\bar{k})\omega_2^2(\bar{k})z^2(\bar{k},t)
\nonumber\\
&-\epsilon x^2(\bar{k},t)+\frac{1}{2}\left[\omega_1^2(\bar{k})+\omega_2^2(\bar{k})\right]\epsilon z^2(\bar{k},t)\bigg{]}\bigg{]}.
\label{12.12}
\end{align}
Finally, to make the path integral converge we take $z(\bar{k},t)$ to be pure imaginary and replace it by the real $y(\bar{k},t)=-iz(\bar{k},t)$, to yield 
\begin{align}
PI(MINK)&=\int D[y]D[x]\exp\bigg{[}\frac{1}{2}\int dt d^3k\bigg{[}i\dot{x}^2(\bar{k},t)-i\left[\omega_1^2(\bar{k})+\omega_2^2(\bar{k}) \right]x^2(\bar{k},t)-i\omega_1^2(\bar{k})\omega_2^2(\bar{k})y^2(\bar{k},t)
\nonumber\\
&-2\epsilon x^2(\bar{k},t)-\left[\omega_1^2(\bar{k})+\omega_2^2(\bar{k})\right]\epsilon y^2(\bar{k},t)\bigg{]}\bigg{]}.
\label{12.13}
\end{align}

To relate this  continuation procedure to the one that we had used for (\ref{12.8}), we note that with $\phi(\bar{x},t)$ being even under parity,  it follows from (\ref{4.5}) that $a_1(\bar{k})=a_1(-\bar{k})$, $a_2(\bar{k})=a_2(-\bar{k})$, $a_1^{\dagger}(\bar{k})=a_1^{\dagger}(-\bar{k})$, $a_2^{\dagger}(\bar{k})=a_2^{\dagger}(-\bar{k})$. Then from (\ref{6.3}) we obtain
\begin{align}
\phi(\bar{x},t)&=(2\pi)^{-3/2}\int d^3kz(\bar{k},t)e^{i\bar{k}\cdot\bar{x}}, \quad \dot{\phi}(\bar{x},t)=(2\pi)^{-3/2}\int d^3kx(\bar{k},t)e^{i\bar{k}\cdot\bar{x}},
\nonumber\\
\ddot{\phi}(\bar{x},t)&=(2\pi)^{-3/2}\int d^3kp_x(\bar{k},t)e^{i\bar{k}\cdot\bar{x}},\quad \dddot{\phi}(\bar{x},t)=(2\pi)^{-3/2}\int d^3k\left[-p_z(\bar{k},t)-\left(\omega_1^2(\bar{k})+\omega_2^2(\bar{k})\right) x(\bar{k})\right]e^{i\bar{k}\cdot\bar{x}}.
\label{12.14}
\end{align}
Substitution of these relations into (\ref{4.1})  then gives (\ref{12.10}), just as it should. Of the functions that appear in (\ref{12.11}) we only need to continue $z(\bar{k},t)$, to completely parallel the continuation of $\phi(\bar{x},t)$ into $\bar{\phi}(\bar{x},t)=(2\pi)^{-3/2}\int d^3ky(\bar{k},t)e^{i\bar{k}\cdot\bar{x}}$.

The utility of (\ref{12.13})  is that  the c-number $y(\bar{k},t)$ and $x(\bar{k},t)$ are the eigenvalues of the self-adjoint and Hermitian q-number $y(\bar{k},t)$ and $x(\bar{k},t)$ operators that appear in the Hamiltonian given in (\ref{7.15}), with  $\vert y(\bar{k},t)\rangle $ and $\vert x(\bar{k},t)\rangle$ being the eigenvectors. In fact because of this direct connection between operators and eigenvalues we can even derive the path integral formula starting from the use of $e^{-i\bar{H}_St}$ as the quantum time evolution operator by putting in intermediate 
$\vert y(\bar{k},t)\rangle $ and $\vert x(\bar{k},t)\rangle$ states at every intermediate time slice, so that the path integral is given by
$\langle i\vert e^{-i\bar{H}_St} \vert f\rangle$ between initial and final states. Thus while we introduced  $y(\bar{k},t)$, $x(\bar{k},t)$, 
 $q(\bar{k},t)$ and $p_x(\bar{k},t)$ in order to address the vacuum normalization issue, we see that they actually provide a very convenient basis  for the scalar field and its derivatives that has additional advantages.
 
 \subsection{The Overall Sign of the Action}
\label{S12D}

In the discussion of the PU oscillator model given in \cite{Bender2008b,Bender2008a} the action was not taken to be that given in (\ref{5.1}), but the slightly more general
\begin{align}
H_{\rm PU}=\frac{\gamma}{2}\int dt\left[{\ddot z}^2-\left(\omega_1^2
+\omega_2^2\right){\dot z}^2+\omega_1^2\omega_2^2z^2\right],
\label{12.15}
\end{align}
where $\gamma$ is a constant. The action in (\ref{5.1}) thus corresponds to $\gamma=1$. Since $\gamma$ is only an overall multiplier the wave equation obtained by varying this action with $z$ and $\dot{z}$ held fixed at the endpoints is not affected, and it still takes the form
\begin{align}
\ddddot{z}+(\omega_1^2+\omega_2^2)\ddot{z}+\omega_1^2\omega_2^2z^2=0.
\label{12.16}
\end{align}
Thus the energy spectrum does not change. However, for arbitrary $\gamma$ the Hamiltonian given in (\ref{5.2}) generalizes  to
\begin{align}  
H_{\rm PU}=\frac{p_x^2}{2\gamma}+p_zx+\frac{\gamma}{2}\left(\omega_1^2+\omega_2^2 \right)x^2-\frac{\gamma}{2}\omega_1^2\omega_2^2z^2,
 \label{12.17}
 \end{align}
to now have an explicit dependence on $\gamma$. The reason why the energy spectrum nonetheless does not change is because (\ref{5.5}), (\ref{5.6}) and (\ref{5.7}) generalize to the $\gamma$-dependent 
\begin{eqnarray}
z(t)&=&a_1e^{-i\omega_1t}+a_1^{\dagger}e^{i\omega_1t}+a_2e^{-i\omega_2t}+a_2^{\dagger}e^{i\omega_2t},
\nonumber\\
p_z(t)&=&i\gamma\omega_1\omega_2^2
[a_1e^{-i\omega_1t}-a_1^{\dagger}e^{i\omega_1t}]+i\gamma\omega_1^2\omega_2[a_2e^{-i\omega_2t}-a_2^{\dagger}e^{i\omega_2t}],
\nonumber\\
x(t)&=&-i\omega_1[a_1e^{-i\omega_1t}-a_1^{\dagger}e^{i\omega_1t}]-i\omega_2[a_2e^{-i\omega_2t}-a_2^{\dagger}e^{i\omega_2t}],
\nonumber\\
p_x(t)&=&-\gamma\omega_1^2 [a_1e^{-i\omega_1t}+a_1^{\dagger}e^{i\omega_1t}]-\gamma\omega_2^2[a_2e^{-i\omega_2t}+a_2^{\dagger}e^{i\omega_2t}],
\label{12.18}
\end{eqnarray}
to yield a Hamiltonian and commutator algebra of the form \cite{Mannheim2000}
\begin{align}
H_{\rm PU}&=2\gamma(\omega_1^2-\omega_2^2)(\omega_1^2 a_1^{\dagger}
a_1-\omega_2^2a_2^{\dagger} a_2)
+\tfrac{1}{2}(\omega_1+\omega_2),
\label{12.19}
\end{align}
\begin{align}
[a_1,a_1^{\dagger}]&=\frac{1}{2\gamma\omega_1(\omega_1^2-\omega_2^2)}, \qquad
[a_2,a_2^{\dagger}]=-\frac{1}{2\gamma\omega_2(\omega_1^2-\omega_2^2)}.
\label{12.20}
\end{align}
Interestingly we note that with negative $\gamma$ but still with $\omega_1^2>\omega_2^2$ the ghost-signatured commutator is moved from the $(a_2,a_2^{\dagger})$ sector to the $(a_1,a_1^{\dagger})$ sector.

In the Hilbert space in which $a_1\vert \Omega \rangle=0$, $a_2\vert \Omega \rangle=0$, it was noted in \cite{Bender2008b,Bender2008a} that even with a $\gamma$ dependence the energy of the vacuum state $\vert \Omega\rangle$ is $(\omega_1+\omega_2)/2$, while the states $a_1^{\dagger}\vert \Omega \rangle$, $a_2^{\dagger}\vert \Omega \rangle$ have energies that respectively lie $\omega_1$ and $\omega_2$ above the ground state energy. Thus despite the explicit presence of  $\gamma$ factors in (\ref{12.19}) and (\ref{12.20}), these factors compensate each other, causing the energy eigenvalues to be independent not just of the magnitude of $\gamma$ but also of its sign. Thus while three out of the four terms in $H_{\rm PU}$ would change sign if the sign of $\gamma$ were to be changed, so would both of the commutators in (\ref{12.20}), doing so in such a way that the energy eigenvalues  would not change sign.

However, there is something that does change substantially, namely the domain of convergence of the wave functions. Specifically,  the ground state wave function given in (\ref{5.4}) generalizes to \cite{Mannheim2007}
\begin{align}
\psi_0(z,x)=\exp\left[\frac{\gamma}{2}(\omega_1+\omega_2)\omega_1\omega_2z^2+i\gamma\omega_1\omega_2zx-\frac{\gamma}{2}(\omega_1+\omega_2)x^2\right].
\label{12.21}
\end{align}
And now for negative $\gamma$ it is the large $x$ behavior  that causes the Dirac norm of the ground state wave function to diverge rather than the large $z$ behavior. Thus now we must continue $x$ and its $p_x$ conjugate into the complex plane rather than $z$ and $p_z$. Thus replacing $x$ by $ -ix=r$ and  $p_x$ by $ ip_x=s$ where $r$ and $s$ are real, the ground state wave function takes the  bounded form
\begin{align}
\psi_0(z,r)&=\exp\left[\frac{\gamma}{2}(\omega_1+\omega_2)\omega_1\omega_2z^2-\gamma\omega_1\omega_2zr+ \frac{\gamma}{2}(\omega_1+\omega_2)r^2\right]
\nonumber\\
&=\exp\left[\frac{\gamma[(\omega_1+\omega_2)r-\omega_1\omega_2z]^2+\gamma\omega_1\omega_2(\omega_1^2+\omega_2^2+\omega_1\omega_2)z^2}{2(\omega_1+\omega_2)}\right],
\label{12.22}
\end{align}
while for operators $r$ and $s$ the Hamiltonian takes the form 
\begin{align}  
\bar{H}_{\rm PU}=-\frac{s^2}{2\gamma}+ip_zr-\frac{\gamma}{2}\left(\omega_1^2+\omega_2^2 \right)r^2-\frac{\gamma}{2}\omega_1^2\omega_2^2z^2.
 \label{12.23}
 \end{align}
 None of these changes affect the energy eigenspectrum.
 
 To construct the Fock space basis, in analog to our analysis of the $\gamma>0$ case described in Sec. \ref{S7B}, and with $\gamma <0$ this time  we replace $(a_1,a_2,a_1^{\dagger},a^{\dagger}_2)$ by $(ia_1,a_2,i\hat{a}_1,\hat{a}_2)$. This yields
\begin{align}
z(t)&=ia_1e^{-i\omega_1t}+a_2e^{-i\omega_2t}+i\hat{a}_1e^{i\omega_1t}+\hat{a}_2
e^{i\omega_2t},
\nonumber\\
r(t)&=-i\omega_1a_1e^{-i\omega_1t}-\omega_2a_2e^{-i\omega_2t}+i\omega_1\hat{a}_1
e^{i\omega_1t}+\omega_2\hat{a}_2e^{i\omega_2t},
\nonumber\\
s(t)&=\gamma[\omega_1^2a_1e^{-i\omega_1t}-i\omega_2^2a_2e^{-i\omega_2t}+\omega_1
^2\hat{a}_1e^{i\omega_1t}-i\omega_2^2\hat{a}_2e^{i\omega_2t},
\nonumber\\
p_z(t)&=\gamma\omega_1\omega_2[-\omega_2a_1e^{-i\omega_1t}+i\omega_1a_2e^{-i
\omega_2t}+\omega_2\hat{a}_1e^{i\omega_1t}-i\omega_1\hat{a}_2e^{i\omega_2t}],
\nonumber\\
a_1e^{-i\omega_1t}&=\frac{1}{2(\omega_1^2-\omega_2^2)}\left[i\omega_2^2z(t)+\frac{s(t)}{\gamma}+i\omega_1r(t)+\frac{p_z(t)}{\gamma\omega_1}\right],
\nonumber\\
\hat{a}_1e^{+i\omega_1t}&=\frac{1}{2(\omega_1^2-\omega_2^2)}\left[i\omega_2^2z(t)+\frac{s(t)}{\gamma}-i\omega_1r(t)-\frac{p_z(t)}{\gamma\omega_1}\right],
\nonumber\\
a_2e^{-i\omega_2t}&=\frac{1}{2(\omega_1^2-\omega_2^2)}\left[\omega_1^2z(t)-\frac{is(t)}{\gamma}+\omega_2r(t)-\frac{ip_z(t)}{\gamma\omega_2}\right],
\nonumber\\
\hat{a}_2e^{+i\omega_2t}&=\frac{1}{2(\omega_1^2-\omega_2^2)}\left[\omega_1^2z(t)-\frac{is(t)}{\gamma}-\omega_2r(t)+\frac{ip_z(t)}{\gamma\omega_2}\right],
\label{12.24}
\end{align}
together with a Hamiltonian and commutation algebra of the form 
\begin{align}
H_{\rm PU}&=-2\gamma(\omega_1^2-\omega_2^2)(\omega_1^2 \hat{a}_1
a_1+\omega_2^2\hat{a}_2 a_2)
+\tfrac{1}{2}(\omega_1+\omega_2),
\label{12.25}
\end{align}
\begin{align}
[a_1,\hat{a}_1]&=-\frac{1}{2\gamma\omega_1(\omega_1^2-\omega_2^2)}, \qquad
[a_2,\hat{a}_2]=-\frac{1}{2\gamma\omega_2(\omega_1^2-\omega_2^2)},
\label{12.26}
\end{align}
so that with $\gamma<0$ the energy eigenspectrum is bounded from below, all commutators are positive, and all norms are positive and finite.

 For the path integral the discussion parallels that given in Sec. \ref{S8}, with (\ref{8.1}) and (\ref{8.2}) being generalized to 
\begin{align}
PI(MINK)=\int D[z]D[dz/dt]\exp\left[\frac{i\gamma}{2}\int_{-\infty}^{\infty} dt\left(\left(\frac{d^2 z}{dt^2}\right)^2-\left(\omega_1^2
+\omega_2^2\right)\left(\frac{d z}{dt}\right)^2+\omega_1^2\omega_2^2z^2\right)\right],
\label{12.27}
\end{align}
and
\begin{align}
PI(MINK,z,x)=\int D[z]D[x]\exp\left[\frac{i\gamma}{2}\int_{-\infty}^{\infty} dt\left(\left(\frac{dx}{dt}\right)^2-\left(\omega_1^2
+\omega_2^2\right)x^2+\omega_1^2\omega_2^2z^2\right)\right].
\label{12.28}
\end{align}
To make the path integral converge we first replace $\omega_1^2$ and $\omega_2^2$ by $\omega_1^2-i\epsilon$ and $\omega_2^2-i\epsilon$. This yields 
\begin{align}
PI(MINK,z,x)=\int D[z]D[x]\exp\left[\frac{\gamma}{2}\int_{-\infty}^{\infty} dt\left(i\left(\frac{d x}{dt}\right)^2-i\left(\omega_1^2
+\omega_2^2\right)x^2+i\omega_1^2\omega_2^2z^2
-2\epsilon x^2+\epsilon\left(\omega_1^2+\omega_2^2\right)z^2
\right)\right],
\label{12.29}
\end{align}
as integrated over paths with real $x$ and real $z$. However, for negative $\gamma$ this time it is the $z^2$ term that is now damped  while the $x^2$ term is not. Consequently, as integrated with a real measure the $\gamma$ negative path integral does not exist, with the path integral thus not existing with a real measure for either sign of $\gamma$. Now the path integral is used to generate time-ordered Green's functions  such as $D(x)=i\langle \Omega\vert T[\phi(x)\phi(0)]\vert\Omega\rangle$. And thus these Green's functions will not be finite, with the vacuum in which the Green's function matrix elements are evaluated thus not being normalizable. Study of the Minkowski path integral thus gives us an alternate way to determine whether or not $\langle \Omega\vert \Omega\rangle$ is finite: the path integral with a real measure either exists or does not exist.

To make (\ref{12.29}) exist we need to damp the $x^2$ term, but not modify the $z^2$ term. Thus we continue $x$ into the complex plane and replace it by $x=ir$, while leaving $z$ real. The path integral for Minkowski time then takes the form 
\begin{align}
PI(MINK,z,r)=\int D[z]D[r]\exp\left[\frac{\gamma}{2}\int_{-\infty}^{\infty} dt\left(-i\left(\frac{d r}{dt}\right)^2+i\left(\omega_1^2
+\omega_2^2\right)r^2+i\omega_1^2\omega_2^2z^2
+2\epsilon r^2+\epsilon\left(\omega_1^2+\omega_2^2\right)z^2
\right)\right].
\label{12.30}
\end{align}
This puts us into a Stokes wedge domain in the complex plane in which the path integral with real $z$ and $r$ and negative $\gamma$ now is fully defined, and now the vacuum state is normalizable. This completely parallels the discussion of $\psi_0(z,r)$ that we gave in (\ref{12.22}) above.

\begin{figure}[htpb]
\centering
\includegraphics[scale=0.5]{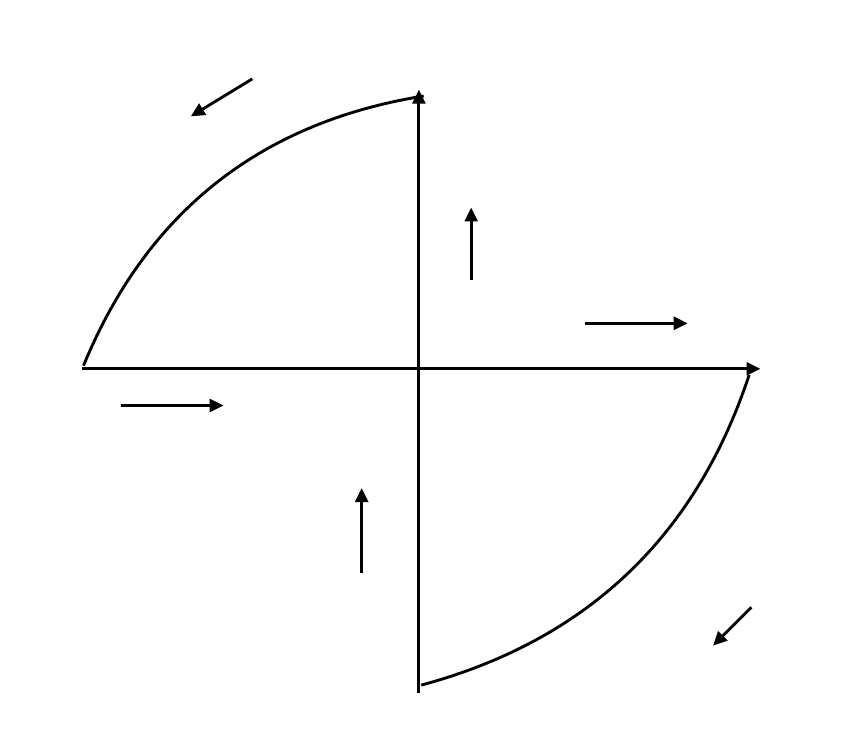}
\caption{Alternate Wick contour}
\label{alternatewickcontour}
\end{figure}

For the Euclidean time path integral approach with $\gamma <0$ we use the Wick contour given in  Fig. \ref{alternatewickcontour}, viz
\begin{align}
\int_{-\infty}^{\infty} +\int_{\infty}^{-i\infty}+ \int_{-i\infty}^{i\infty}+ \int_{i\infty}^{-\infty}={\rm pole~terms~plus ~cut~contributions},
\label{12.31}
\end{align}
i.e., along the real axis, then lower-half-plane quarter circle, then up the imaginary axis, and then upper-half-plane quarter circle.  Assuming no pole, cut  or circle contributions, and on setting $\tau=it$ and letting $I$ denote the action,  from (\ref{12.27}) and (\ref{12.28}) we obtain 
\begin{align}
I(MINK,z,x)&\equiv \int_{-\infty}^{\infty}idt\equiv -\int_{-i\infty}^{i\infty}idt=+\int_{-\infty}^{\infty} d\tau \equiv I(EUCL,z,x),
\nonumber\\
PI(EUCL,z,x)&=\int D[z]D[dz/d\tau]\exp\left[\frac{\gamma}{2}\int_{-\infty}^{\infty} d\tau\left(\left(\frac{d^2 z}{d\tau^2}\right)^2+\left(\omega_1^2
+\omega_2^2\right)\left(\frac{d z}{d\tau}\right)^2+\omega_1^2\omega_2^2z^2\right)\right]
\nonumber\\
&=\int D[z]D[x]\exp\left[\frac{\gamma}{2}\int_{-\infty}^{\infty} d\tau\left(\left(\frac{dx}{d\tau}\right)^2+\left(\omega_1^2
+\omega_2^2\right)x^2+\omega_1^2\omega_2^2z^2\right)\right]. 
\label{12.32}
\end{align}
Given that we are taking $\gamma$ to be negative this time, we see that with real $z$ and real $x=dz/d\tau$ the Euclidean path integral is well behaved on every path. (The same is true of the analog relativistic second-order plus fourth-order scalar field theory path integral  \cite{Hawking2002}.) However, the Minkowski time path integral with a real measure is not. Thus we conclude that the pole and/or  cut and/or circle contributions are not only not ignorable, they generate an infinite contribution. Hence their contribution in a Wick rotation cannot be ignored and the Euclidean time path integral does not correctly describe the situation. 

In parallel, if we set $x=ir$, then (\ref{12.32}) is replaced by
\begin{align}
I(MINK,z,r)&\equiv I(EUCL,z,r),
\nonumber\\
PI(EUCL,z,r)&=\int D[z]D[r]\exp\left[\frac{\gamma}{2}\int_{-\infty}^{\infty} d\tau\left(-\left(\frac{dr}{d\tau}\right)^2-\left(\omega_1^2
+\omega_2^2\right)r^2+\omega_1^2\omega_2^2z^2\right)\right].
\label{12.33}
\end{align}
And since the $z$ and $r$ path integrations are independent, now it is the Euclidean time path integral that is not well defined when $z$ and $r$ are both real. Thus with either real $x$ or real $r=-ix$, in neither case are the Minkowski time and Euclidean time path integrals simultaneously finite.

\section{Final Comments}
\label{S13}
For a  quantum field theory to be physically relevant it must be formulatable in a Hilbert space with an inner product that is time independent,  finite and positive (though zero norm is also acceptable). However, in and of itself, specifying an action and a set of canonical commutators is not enough to either fix the Hilbert space or specify the appropriate inner product. Ordinarily, one supplements these requirements with the additional (generally regarded as self-evident) requirements that the fields and the Hamiltonian of the theory be Hermitian, and that the inner product be the standard, presumed finite,  Dirac $\langle n\vert n \rangle$ one. (In the axiomatic approach to quantum field theory   \cite{Streater1964} the finiteness of $\langle \Omega \vert \Omega \rangle$ is one of the starting assumptions.) However, it is not automatic for an arbitrary theory that $\langle \Omega \vert \Omega \rangle$ be finite, and  so one needs to check on a case by case basis. And in this paper we have presented a procedure for doing so. The procedure is based on using the occupation number space representation  to construct an equivalent wave mechanics representation, from which we can then check for the normalizability of the vacuum state, and accordingly of the states that can be excited out of it. An alternative but equivalent approach is to check whether or not the Minkowski path integral with a real measure exists. If it does not, then the standard Dirac inner product is not finite. 

Using the occupation number space representation  procedure we have found a case, a second-order plus fourth-order  scalar field theory, in which the standard Dirac inner product $\langle n\vert n \rangle$ actually is not finite. In this example  the Minkowski time path integral with a real measure diverges even though the Euclidean time path integral does not. In this case contributions from the Wick rotation contour cannot be ignored, and the use of a Euclidean time path integral can be misleading. Thus even if a Euclidean time path integral is well behaved, it only gives a good description of the theory if the Minkowski time path integral is well behaved too. Since $\langle \Omega\vert \Omega\rangle$ is not finite for the second-order plus fourth-order  scalar field theory, use of the standard Feynman rules is not valid, with these rules not only leading  to states with negative norm, they lead to states with infinite negative norm. This lack of finiteness means that the Hamiltonian is not self-adjoint when acting on these particular states.

However, the Hamiltonian of the second-order plus fourth-order  scalar field theory is $PT$ symmetric, so we can use the techniques of the $PT$-symmetry program and continue the fields and the Hamiltonian in this theory into the complex plane. There is then a domain in the complex plane in which one can define an appropriate time-independent, positive and finite inner product, viz. the  $\langle L\vert R\rangle$ overlap of left-eigenstates and right-eigenstates of the resulting Hamiltonian, with the resulting vacuum state then being normalizable, and with there being no states with negative or infinite $\langle L\vert R\rangle$ norm \cite{footnote5}. In this domain it is the Euclidean time path integral that diverges while the Minkowski time path integral does not. So again there are contributions from the Wick rotation contour. In this domain the second-order plus fourth-order  scalar field theory is fully consistent, unitary and renormalizable, with this analysis being relevant to the construction of a consistent, unitary and renormalizable quantum theory of gravity, a theory that can serve as an ultraviolet completion of Einstein gravity \cite{footnote7}. And even though there is a continuation of the fields into the complex plane, we have shown that the resulting classical limit of the theory is nonetheless completely real.

\begin{acknowledgments}
The author wishes to thank Dr. A. Barvinsky, Dr. C. M. Bender, Dr.  J. Feinberg and Dr. O. Lechtenfeld for helpful comments.
\end{acknowledgments}


\begin{thebibliography}{99}
\bibitem{footnote1} The difference in sign between the fourth-order (\ref{4.4}) and the second-order (\ref{1.7}) is due to the fact that $[\phi,\dddot{\phi}]$ and $[\phi,\dot{\phi}]$ as respectively given in  (\ref{4.3}) and (\ref{1.2}) have opposite signs.
\bibitem{footnote2}  If $AHA^{-1}=H$, where $A$ is a general antilinear operator, where $H$ is a general linear operator, and where $H\vert \psi\rangle=E\vert \psi\rangle$, then 
 $AH\vert \psi\rangle=AE\vert \psi\rangle=E^*A\vert \psi\rangle=AHA^{-1}A\vert \psi\rangle=HA\vert \psi\rangle$. Thus for every eigenvalue $E$ there is an eigenvalue $E^*$. If $A\vert \psi\rangle$ is equal to $\vert \psi\rangle$ then the energies are real, though in and of itself this analysis does not guarantee that every $E$ is positive. If  $A\vert \psi\rangle$ is not equal to $\vert \psi\rangle$ then the energies are in a complex pair. Both $H_{\rm PU}$ and $H_{S}$  are  in the $E=E^*$ realization in which  all energies are real.  While this analysis would apply to any antilinear symmetry, since it was first discovered in the $PT$ context, even in the general case  theories with antilinear symmetry are referred to  as  $PT$ theories. As discussed in \cite{mannheim2018antilinearity}, under the requirements solely of complex Lorentz invariance and  probability conservation one can show that a theory must be $CPT$ invariant, where $C$ denotes charge conjugation. For the theories discussed in this paper $C$ is separately conserved, with  $CPT$ then defaulting to $PT$.  For completeness we note that if  $AHA^{-1}=-H$, where $A$ is a general antilinear operator, where $H$ is linear, and where $H\vert \psi\rangle=E\vert \psi\rangle$, then 
 $AH\vert \psi\rangle=AE\vert \psi\rangle=E^*A\vert \psi\rangle=AHA^{-1}A\vert \psi\rangle=-HA\vert \psi\rangle$. If $A\vert \psi\rangle$ is equal to $\vert \psi\rangle$ then $E=-E^*$ and the energies are pure imaginary.   If  $A\vert \psi\rangle$ is not equal to $\vert \psi\rangle$  then $E$ and $-E^*$ are separate energy eigenvalues. However, this set could include real $E$, in which case the energies are real $E$ and real $-E$. This is actually the case for the $PT$ odd position operator as its eigenvalues range from $-\infty$ to $+\infty$ with $PT\vert x\rangle=\vert -x\rangle$.
\end{thebibliography}
\end{document}